\documentclass[twocolumn,aps,prb, floatfix,superscriptaddress,nofootinbib, sortedaddress]{revtex4-2}

\usepackage{graphicx, color, soul}
\usepackage[usenames,dvipsnames, pdftex]{xcolor}
\usepackage{dcolumn}
\usepackage{bm}
\usepackage{amsmath, amsfonts, amssymb, mathrsfs, dsfont, mathtools}
\usepackage[mathscr]{eucal}
\usepackage{textcomp}
\usepackage{appendix}
\usepackage{physics}

\usepackage[colorlinks, breaklinks, 
            linkcolor=Blue,
            citecolor=Blue,
            urlcolor=Blue]{hyperref}
            
\usepackage[all]{hypcap} 
\usepackage{enumitem}
\usepackage[normalem]{ulem} 

\newcommand{\Le}{\left}
\newcommand{\Ri}{\right}
\newcommand{\nn}{\nonumber}
\newcommand{\f}{\frac}

\newcommand{\ra}{\rangle}
\newcommand{\la}{\langle}
\newcommand{\eq}[1]{\begin{align}#1\end{align}}

\newcommand{\msr}{\mathscr}

\newcommand{\SB}[1]{{\color{Black} #1}}

\newcommand{\mSP}{{{\mathscr S}}_\text{P}} 

\newcommand{\mSE}{{{\mathscr S}}_\text{E}}
\newcommand{\Sd}{{{\mathscr S}}_\text{d}}
 
\newcommand{\Ga}{\Gamma} 
\newcommand{\bej}{\beta_j(n)}

\begin{document}

\title{Inhomogeneous Floquet thermalization}

\author{Soumya Bera}
\affiliation{Department of Physics, Indian Institute of Technology Bombay, Mumbai 400076, India}

\author{Ishita Modak}
\affiliation{Department of Physics, Indian Institute of Technology Bombay, Mumbai 400076, India}

\author{Roderich Moessner}
\affiliation{Max-Planck-Institut f\"ur Physik komplexer Systeme, 01187 Dresden, Germany}

\date{\today}

\begin{abstract}
How a closed system thermalizes, especially in the absence of global conservation laws but in the presence of disorder and interactions, is one of the central questions in non-equilibrium statistical mechanics. 
We explore this for a disordered, periodically driven Ising chain. 
Our numerical results reveal inhomogeneous thermalization leading to a distribution of thermalization timescales within a single disordered sample, {which we encode via a distribution of effective local  temperatures.} 
Using this, we find an excellent collapse {\it without any fitting parameters} of the local {relaxation dynamics} for the entire range of disorder values in the ergodic regime when adapting the disorder-averaged diagonal entanglement entropy as internal `time' of the system. 
This approach evidences a remarkably uniform parametrization of the dynamical many-body evolution of local temperature 
within the otherwise highly heterogeneous ergodic regime, independent of the strength of the disorder.
\end{abstract}

\maketitle

\section{Introduction}
Spatially heterogeneous relaxation dynamics towards equilibrium is a hallmark of nonergodicity, being found in paradigmatic settings of glasses and jammed systems~\cite{BerthierRMP11}. 
Such dynamical heterogeneity, e.g., evidenced in the coexistence of different relaxation timescales,  can arise from spatial variations associated with the presence of metastable states.  
This can lead to global nonexponential decay of correlation functions in time despite local exponential decay rates. 
Alternatively, relaxation processes can be inherently complex also, contributing to local nonexponential decay~\cite{GlotzerPRE98}.
In ergodic systems, such as supercooled liquids, even regions of slower relaxation eventually thermalize~\cite{EdigerARPC00}.

In quantum systems undergoing unitary dynamics, the nature of thermalization, or its absence, has been a focal point of research in recent years~\cite{Nandkishore2015, DAlessio2016, UedaNat2020}. 
Thermalization in ergodic systems occurs, loosely speaking, as sub-system density matrices evolve to a thermal state, with the remainder of the system effectively acting as its bath.
The presence of disorder may impede thermalization, for instance, by the emergence of a set of quasi-local integrals of motion, resulting in emergent effective integrability and non-ergodicity, a phenomenon dubbed many-body localization~(MBL)~\cite{Gornyi2005,  Basko2006, Nandkishore2015, Bera2015, AbaninBloch-Review-2018, AletReview2018, Sierant_2025}. 
In this parameter regime, spatial and dynamical heterogeneity has been observed in local entanglement measures~\cite{BeraArul16, ClaudiaPRB22} as demonstrated in models such as the XXZ spin chain.
Indeed, entanglement can exhibit substantial spatial heterogeneity in this regime, as indicated by the subvolume scaling of the standard deviation of the cut-to-cut entanglement entropy~\cite {Khemani2017}. 
Because of the difficulty of studying the real-time dynamics of disordered interacting quantum systems, a definite consensus on the nature of this regime has been slow to emerge, e.g., Refs.~\cite{Weiner19, PandaMBL19, SirkerPRL20, Luitz2020, SuntasPRB20, AbaninAOP21, Polkovnikov2021, SelsBathPRB22, SierantPRB22, Morningstar2022, CrowleySciP22, LongPRL23,  EversPRB23, ChavezUltraSlow23}.

Here, we show that even the weakly disordered {\it ergodic} regime can exhibit considerable spatial structure, which we investigate in detail. We focus on a nonintegrable driven disordered Ising chain without global conservation laws.
While even weak disorder tends to slow down relaxation, eventual thermalization can be remarkably robust~\cite{Weiner19}. 
We analyze thermalization after a sudden quench via the time evolution of the local subsystem (inverse) temperature, $\bej$, where $j$ denotes a bond involving two sites after $n$ the time steps.  
The $\bej$ evolution reveals apparent {glassiness} in the sense of a distribution of local relaxation timescales within a single disorder configuration. 
Indeed, the time evolution of the spatial and disorder averaged $\bej$ shows well-developed nonexponential behavior in finite-size numerical simulation. 
With increasing disorder, we further observe mixed dynamics, i.e., locally nonexponential decay of the $\bej$ accompanied by regions in space with exponential relaxation in time towards `infinite temperature'.

The broad distribution of thermalization times, even at a single disorder value, suggests that relying on a single time scale may not be sufficient. 
Relatedly, it is challenging to compare the time evolution between different disorder realizations and, further, strengths.  Here, we propose a unified description employing the ensemble-averaged diagonal entanglement entropy $\Sd(t)$ to monitor the dynamical evolution of $\bej$ across the entire ergodic regime. This 
extends the framework introduced by \textcite{EversPRB23} for non-driven systems, which used the average entanglement entropy to track particle density imbalance decay.
We find a straightforward data collapse of the evolution of average inverse temperature for several disorder values, offering a homogeneous perspective on thermalization dynamics despite its inherent heterogeneity.
While previous work~\cite{EversPRB23} required extra fitting parameters to obtain a scaling collapse,  our intrinsic parametrization of time obviates this need.

\section{Driven Ising Model}
The time-dependent Hamiltonian of this periodically driven system is defined as, 
\eq{
    H(t) = \begin{cases}
        2H_x & \text{if } 0 < t < \frac{T}{2} \\
        2H_z  & \text{if } \frac{T}{2} < t < T, \nn
    \end{cases}
}
\eq{
\label{Floquet_hamiltonian}
    & H_x = \sum_{i=1}^{L} g\Gamma\sigma_i^x, \nn \\
    & H_z = J \sum_{i=1}^{L-1} \sigma_i^z \sigma_{i+1}^z + \sum_{i=1}^L (h+g\sqrt{1-\Gamma^2}G_i)\sigma_i^z, 
}
$\sigma_i^x$ and $\sigma_i^z$ are the Pauli matrices on site $i$. 
We follow the standard parametrization of the model, see Refs.~\cite{ZhangFloquetPRB16, TaliaPRB19, SierantPRBising23}.
The interaction strength $J=1$ and $g, h $ and $T$ are 0.9045, 0.8090 and 0.8, respectively. 
Such a parametrization is motivated by the clean static model, where strong thermalization is obtained with these parameter values for system sizes readily accessible in exact diagonalization studies~\cite{KimPRL13}. 
The longitudinal field is disordered and chosen from a Gaussian distributed random variable $G_i$ with zero mean and unit variance. The $\Gamma$ controls the disorder strength, and the model is believed to have an MBL transition at $\Gamma \simeq 0.3$~\cite{ZhangFloquetPRB16, TaliaPRB19}. 
The stroboscopic time evolution is performed with Floquet operator $U_\text{F} (T) = e^{-i H_x T/2} e^{-i H_z T/2}$ using a Hadamard transformation, see Ref.~\cite{TaliaPRB19} for further details. 
The initial state for these calculations is the N\'eel state $|1010\ldots\rangle$. 
%

\begin{figure}[!t]
    \centering
    \includegraphics[width=1.0\columnwidth]{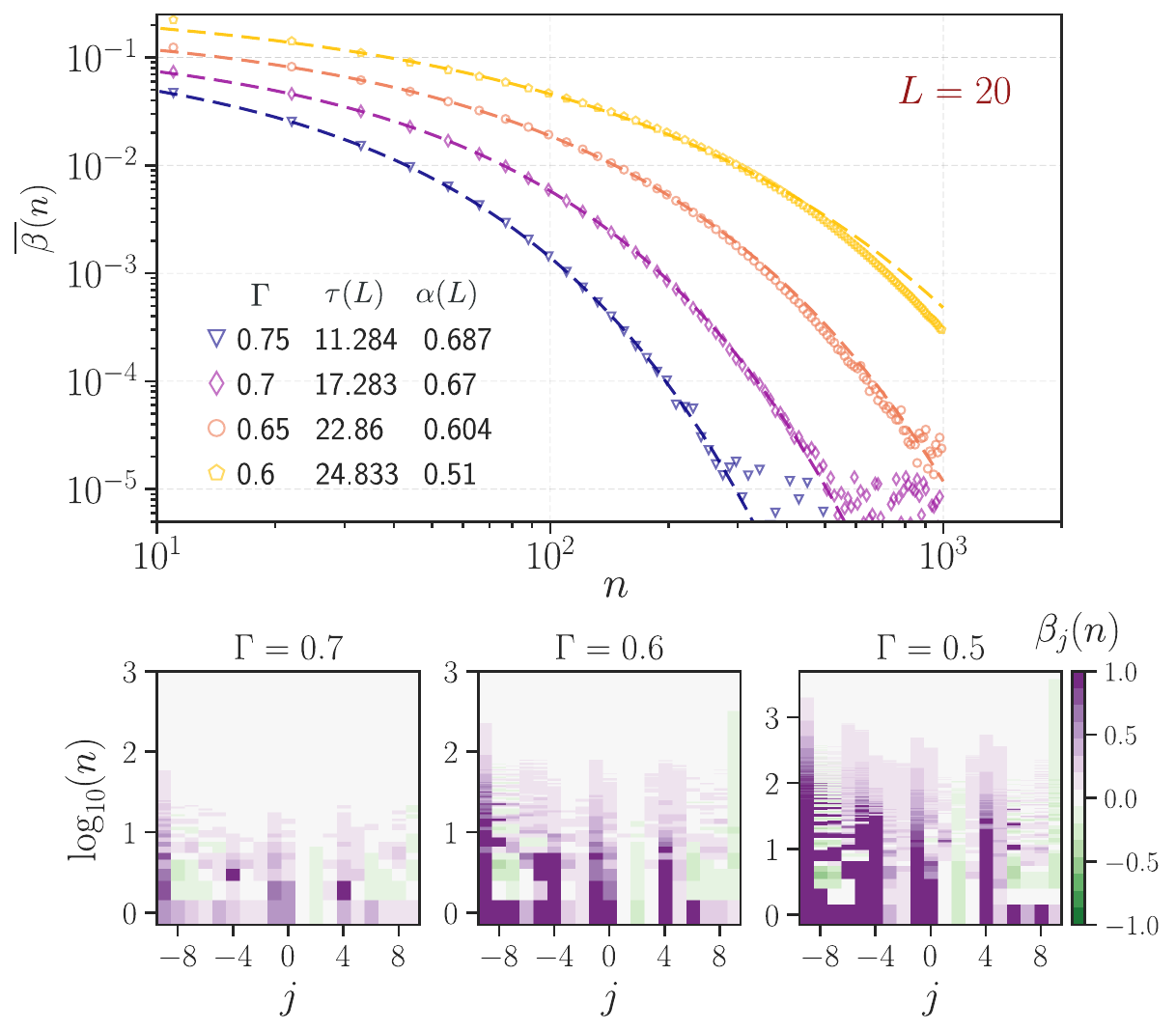}
    \caption{Stretched exponential decay of the average local temperature $\overline{\beta}(n)$ with Floquet time step $n$. The line indicates a stretched exponent fit: $A \exp[-(t/\tau)^\alpha)]$. The corresponding values of the thermalization time $\tau(L)$ and the stretched exponent $\alpha(L)$ are provided in the legend for $L=20$. Lower panel: Spatiotemporal inhomogeneity of the evolution of local bond temperature  $\beta_j(n)$ as defined in Eq.~\eqref{eq:beta} for different values of $\Ga={0.7, 0.6, 0.5}$ as a function of $\log(n)$ for a typical disorder configuration. The white coloring represents infinite temperature ($\beta_j{=}0$). 
    }
    \label{fig:decay_locT_map}
\end{figure}
\begin{figure*}[!tb]
    \centering
    \includegraphics[width=1\textwidth]{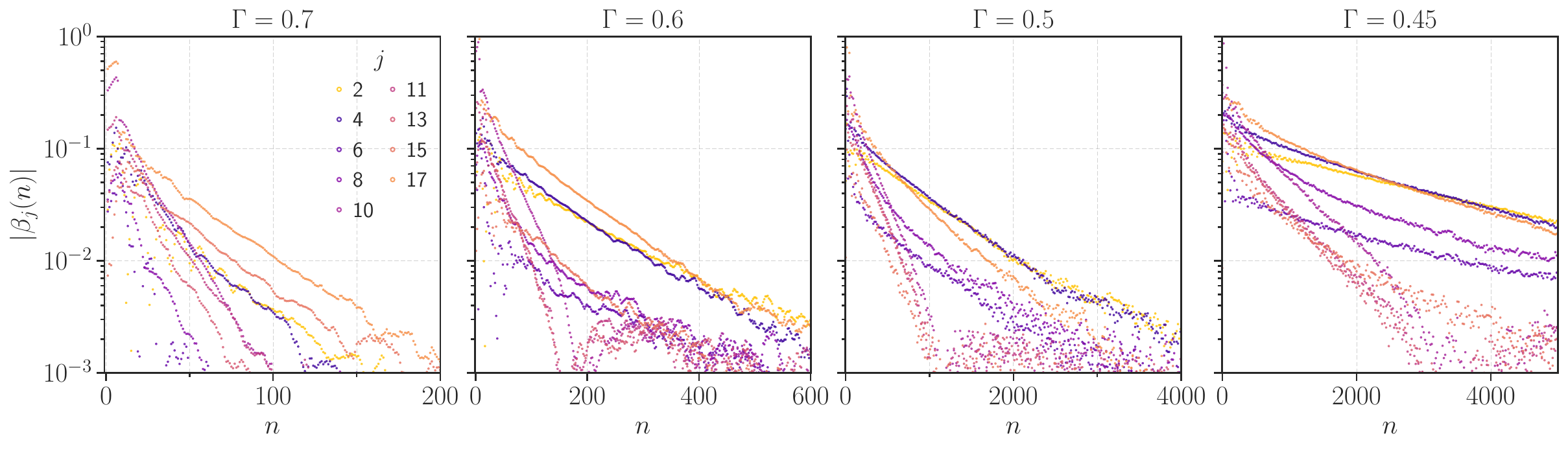}
    \caption{Time evolution of local inverse temperature $|\beta_j(n)|$ at different locations, $j$, for several disorder 
    {values $\Gamma=0.6,0.5, 0.4, 0.45$ for $L=20$. Log-linear scale highlights exponential decay at large $\Gamma$ for a typical disorder configuration. With decreasing $\Gamma$, the thermalization time increases.  For the small value of $\Gamma =0.45$, the curves \SB{for this sample} appear to fall into two groups characterized by rapid and slow decay rates. }}
    \label{fig:beta_samples}
\end{figure*}
\section{Observables}
\subsection{Local temperature}
The local temperature $\beta_j(n)$ is defined  for each bond by minimizing the Frobenius norm distance between the canonical density matrix $e^{-\beta_j \msr{H}_\text{b}}$, and the sub-system density matrix $\rho_j^\text{b}$, 
\eq{
\text{min}_{\beta_j} \, \Le \lvert e^{-\beta_j \msr{H}_\text{b}} - \rho_{j}^\text{b}(n) \Ri \rvert \ ,
\label{eq:beta}
}
with $\msr{H}_\text{b}(j) = \sigma^z_j \sigma^z_{j+1} + h_j/2 \, \sigma^z_j \otimes \mathds{1}_{2\times 2} +  h_{j+1}/2 \,  \mathds{1}_{2\times 2} \otimes  \sigma^z_j  + \{z \leftrightarrow x, h_j \leftrightarrow g_j \}$ is the local bond Hamiltonian with $h_j, g_j$ being the local fields, and $\rho^\text{b}_j(n) = \text{Tr}_{L-\{j,j+1\}} \, |\psi(n) \ra \la \psi(n) |$ is the reduced density matrix for that bond. 
The norm is defined as $\lvert A \rvert \equiv \text{tr}(\sqrt{A^{\dagger} A})$. 
We expand the definition of $\beta$ from Ref.~\cite{BurkeHaquePRE23} to the time domain. 
This approach includes the time-evolved wavefunction's structure through the sub-system density matrix.
It is important to note that the precise value of $\beta_j(n)$ is contingent upon the chosen definition of norm, as examined in detail in Ref.~\cite{BurkeHaquePRE23}.

\section{Results}
\subsection{(Non-)Exponential Heating}
An interacting driven system heats to an infinite temperature featureless state~\cite{LazaridesPRL14, DalessioPRX14}.
The heating rate $\tau$ generally depends exponentially on the drive frequency,  $\tau \propto \exp(\omega/J)$ for $\omega/J \gg 1$, where $J$ is a microscopic energy scale~\cite{AbaninPRL15, MoriPRL16, AbaninPRB17, MallayyaPRL19}.
In the presence of disorder, it has been seen that the (average) heating slows down considerably~\cite{PontePRL14, LazaridesMBLPRL15,  RehnPRB16, BordiaNatPhy2017, TaliaPRB19}.
As shown in Fig.~\ref{fig:decay_locT_map}, the time evolution of the disorder and spatially averaged (indicated by the overline) $\overline{\beta}(n)$ indeed exhibits a stretched exponential decay, with the decay exponent inversely correlated to the strength of the disorder,  similar to the decay of the correlation function as reported earlier in Refs.~\cite{TaliaPRB19, LongPRL23, Asmi23}. A pronounced finite size effect is also observed and shown in App.~\ref{app:fs}.

The lower panel of Fig.~\ref{fig:decay_locT_map} highlights the spatially inhomogeneous evolution of $\bej$ for a given disorder configuration for several $\Ga$ values. 
\SB{In a given sample, one would expect thermalization time to fluctuate due to disorder fluctuations between different sites. 
In particular, small fluctuations in disorder would imply exponential sensitivity to decay time, provided the thermalization is exponentially fast at all bonds in the Floquet drive cycle $n$. }
The upper panel of Fig.~\ref{fig:beta_samples} shows $\beta_j(n)$ at different locations for a single disorder configuration.  
For weak disorder, $\Gamma=0.7$, thermalisation is exponential everywhere, $e^{-n/\tau_j}$,  however with spatially varying $\tau_j$. 
The $| \bej |$ decays to  ${\sim} 10^{-3}$ for $L=20$, and is expected to vanish in the thermodynamic limit. 
With increasing disorder, the variation in $\tau_j$ increases, and for even stronger disorder $\Gamma=0.45$, the heating becomes slow, possibly as a stretched exponential (see Fig.~\ref{fig:beta_samples}(c)). 
Even there, a few subsystems still exhibit fast thermalization, i.e., an exponential decay of $|\bej|$ in time, $n$. 
This distribution of thermalization time scale is reflected in a stretched exponential decay, $e^{-(t/\tau)^\alpha} = \int du P(u) e^{(-t/u)}$,  of the average $\overline{\beta}(n)$ as highlighted in Fig.~\ref{fig:decay_locT_map} with a timescale that increases, and an exponent $\alpha$ which decreases, with increasing disorder. 

\subsection{Thermalization time}
Fitting each $\bej$ trace is impractical because of fluctuations in the data and uncertainties associated with the fit. 
We instead extract a local decay time $\tau_j$ via 
\eq{
\tau_j \coloneqq \f{\int_0^T n |\bej| dn} {\int_0^T |\bej| dn}.
\label{eq:tau}
}
Concretely, the pure exponential $|\bej| \propto e^{-n/\tau_j^\prime}$ yields $\tau_j=\tau_j^\prime + T/(1+e^{T/\tau_j^\prime})\rightarrow \tau_j^\prime$ for sufficiently large number of  Floquet time steps $T$.
%
Indeed, for the larger $\Ga =0.7$, the individual samples show exponential decay within the simulation time window.
With decreasing $\Ga< 0.5$, not all the traces of $\bej$ show pure exponential decay; instead, there is mixing of both exponential and stretched exponential decays (Fig.~\ref{fig:beta_samples}, right panel). 
This affects the calculation of the decay time as defined in Eq.~\eqref{eq:tau}, and we refrain from doing this analysis for smaller values of the $\Ga$, i.e., larger disorder values close to the putative MBL transition. 
In this regime, the simple way of describing the exponential heating dynamics using the Fermi-Golden rule~\cite{MallayyaPRL19, TatsuhikoPRB21, RakcheevPRR22} is probably inapplicable, and one might need to go beyond this perturbative treatment.  
\begin{figure}[!b]
    \centering
    \includegraphics[width=1\columnwidth]{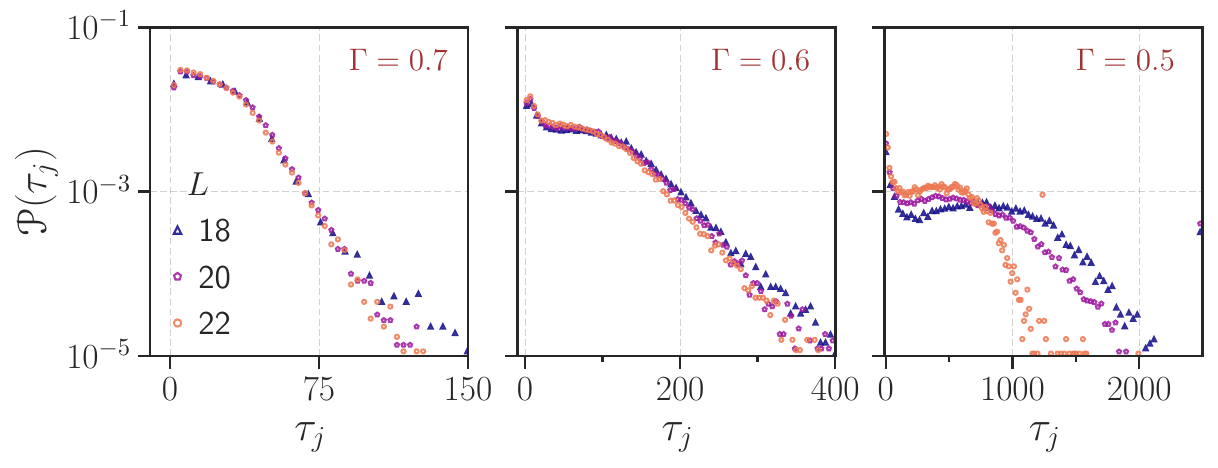}
    \caption{Probability distribution of the local thermalization time $\tau_j$ for system sizes $L=18, 20, 22$, and disorder values $\Ga=0.7, 0.6, 0.5$.  The distribution is broad with an exponential tail. For smaller $\Gamma=0.5$, the distribution shows strong finite-size effects. Typically, $\sim2.5 {\times} 10^3$ disorder configurations are used for these distributions. }
    \label{fig:beta_tau_dist}
\end{figure}

%
Figure~\ref{fig:beta_tau_dist} shows the $\tau_j$ distribution in the ergodic phase for $\Ga=0.7, 0.6, 0.5$.  
A pronounced exponential tail is observed for all the disorder values, with a plateau forming at smaller $\Ga$ - \SB{we associate this plateau feature in the distribution function with the broad distribution of timescales.} 
At $\Gamma=0.5$, significant finite size and time effects are evident (see in App.~\ref{app:fs}), suggesting the need for larger system sizes $L\gtrsim 22$ to observe thermalization in all parts of the sample.  

For the largest system $L=22$, the $\msr{P}(\tau_j)$ shifts towards the origin compared to smaller system sizes, indicating a slow flow of the full distribution towards shorter thermalization time.

\subsection{Density matrix evolution}
\begin{figure}[!tb]
    \centering
\includegraphics[width=0.99\columnwidth]{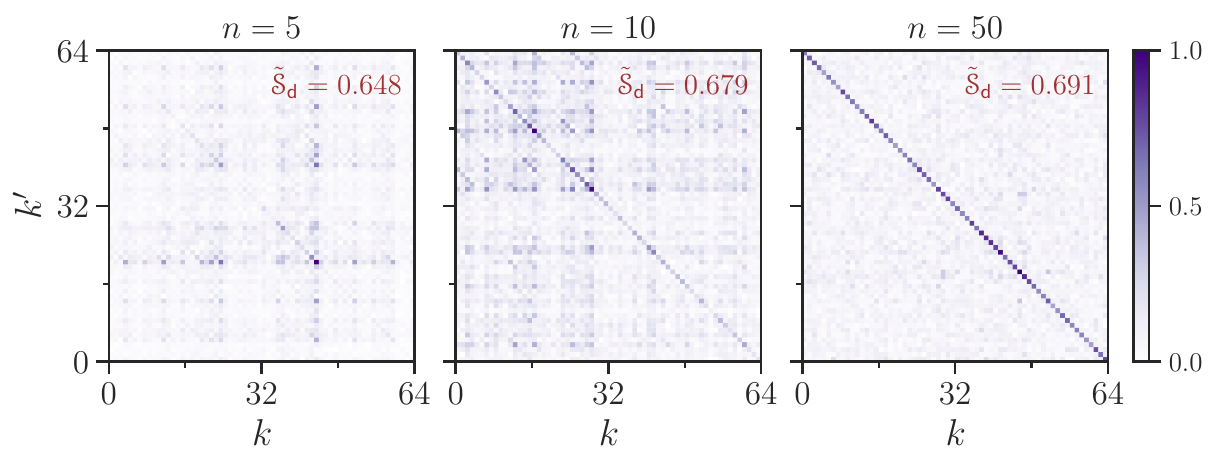}
    \caption{Time evolution of the half-system density matrix $\rho^\text{A}_{k k^\prime}$ (normalized such that its maximum value is unity) for $\Ga=0.7$, $L=12$ and for a typical disorder configuration. As expected for a featureless random state, the diagonal elements dominate over time as the system heats up. Additionally, the diagonal entropy reaches maximum  $\tilde{\msr{S}}_\text{d}=\msr{S}_\text{d}/(L/2) \simeq \log(2)$,  as indicated in the figure.}
    \label{fig:density_map}
\end{figure}

Having established a relatively broad distribution of $\tau_j$,  resulting in a stretched exponential decay of the spatially averaged $\bej$ (Fig.~\ref{fig:decay_locT_map}), we now provide an entanglement perspective to the thermalization process.
Figure~\ref{fig:density_map} shows a typical evolution of the half-system density matrix $|\rho_{n n^\prime}|/\text{max}(\rho_{n n^\prime})$ for different evolution times, $n$,  for $\Ga = 0.7$, which becomes thermal at these times. 
This behavior is representative of all samples reaching thermal equilibrium. 
\subsection{Diagonal entanglement entropy}
The entanglement entropy is defined as, 
$
\mSE = -\text{Tr} \left( \rho^\text{A} \log \rho^\text{A} \right), 
$
where $\rho^\text{A}$ is the reduced half-system density matrix. 
At long time, for the ergodic system, $\mSE$ reaches the Page value i.e., $\mSE = L/2 \log(2)-1/2$~\cite{Page93}. 
\SB{For such state}, the diagonal elements {$\rho_{kk}^A$} dominate; they scale as $\rho_{kk}^A\propto 1/\sqrt{\msr{D}},$ while the off-diagonal terms are suppressed as $\sqrt{\msr{D}^{1/2}}$~\cite{ArulPRE01}, where $\msr{D}=2^{L}$ is the Fock space dimension. 
\SB{The diagonal entropy in this basis is defined as} 
\eq{
\msr{S}_\text{d} = -\sum_k\rho_{kk}^A \log \rho_{kk}^A.
\label{eq:sdiag}
}
For a Haar random state, $\rho_{kk}^A\propto 2^{-L/2}$, therefore $\msr{S}_\text{d} \propto (L/2) \log (2)$, the volume law scaling for ergodic systems.
Further, the diagonal entropy can be expressed in terms of the participation entropy $\msr{S}_\mathrm{P} = 2 \msr{S}_\text{d}$~(see App.~\ref{app:sd} for details for the driven Ising model), where $
\mSP(n) = -\sum_{j=1}^{\msr{D}} p_j(n) \log p_j(n) 
$, with $p_j(n) = |\la j | \psi(n)\ra|^2$ is the probability of occupation of each spin basis state $|j\ra$, and $|\psi(n)\ra$ is the time evolved wavefunction.

Therefore, $\Sd$ is an alternative measure of delocalization in configuration space. 
Indeed, Ref.~\cite{DetomasiPRL20} showed that for a pure state with few random non-zero elements relative to the dimension of the space, the scaling of  $\Sd$ with subsystem volume $L/2$ is exactly given by that of its participation entropy $\mSP$.

\begin{figure}[!tb]
    \centering
    \includegraphics[width=1\columnwidth]{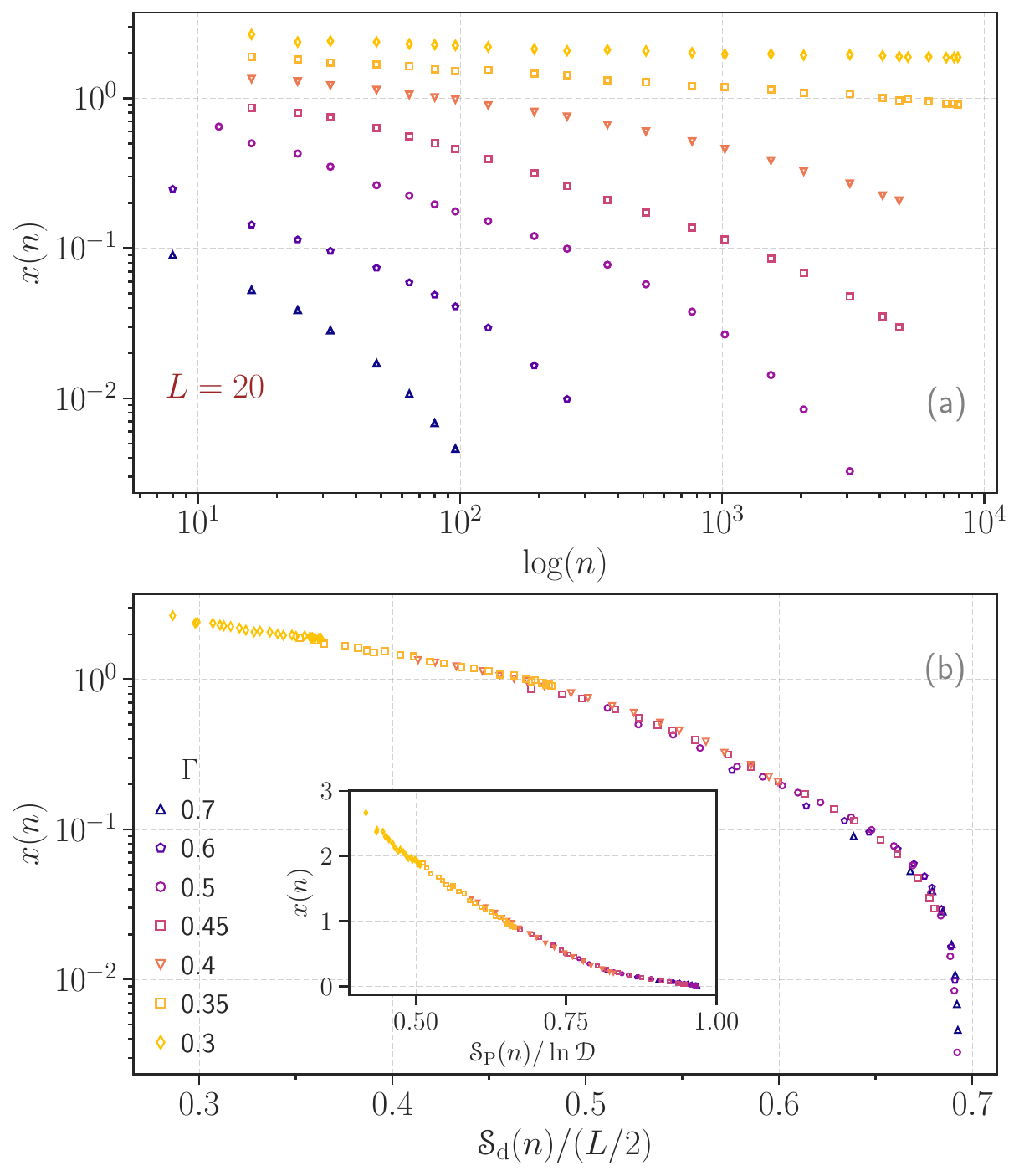}
    \caption{Time evolution of the disorder averaged typical inverse local temperature $x(n) = \la \overline{\beta(n)}\ra_\text{typ.}$ for several disorder values until the putative transition point $\Ga_\text{c} \simeq0.3$ and for $L=20$. The initial slow growth gives way to faster thermalization at later $n$. This feature has yet to manifest for smaller $\Gamma$. Below: The collapse of the evolution when diagonal entropy $\Sd$ is adapted as an implicit measure of time, $n$, for the above figure. Inset: The same data, plotted against the $\mSP$, normalized with the Fock space dimension $\mathscr{D}$.}
    \label{fig:beta_collapse}
\end{figure}
%
\subsection{Synchronized dynamics}
The upper panel of Fig.~\ref{fig:beta_collapse} shows the time evolution of the disorder averaged \SB{$x(n) = \la \overline{\beta(n)}\ra_\text{typ.}$} with Floquet cycle $n$ for several values of disorder  $\Ga=0.7 \ldots 0.3$. \SB{The overline denotes the average over the bonds $j$, and $\la . \ra_\text{typ.}$ denotes the median value of $\bej$ across different disorder configurations}.
For intermediate disorder values, $x(n)$ shows slow propagation (see, e.g., $\Ga=0.45$); however, with increasing time, it accelerates as seen by the rightward bending of the curve to eventual thermalization.  
Such bending happens at progressively higher $n$ with increasing disorder, and to observe this at an even larger disorder, larger $L$ and $n$ are necessary.

Most strikingly,  when $\Sd$ is adapted as an ensemble average internal time of the system, we observe an excellent collapse of the mean $x(n)$ for several disorder values as seen in Fig.~\ref{fig:beta_collapse}. 
This collapse, requiring no fitting parameters, implies that the diagonal entropy $\Sd$ faithfully describes central aspects of the thermalization of the closed system, such as the time to heat up to infinite temperature at finite disorder values.
Concretely, disorder slows down entropy production, thus delaying thermalization. 
Once the simulation time is parametrized by the entropy itself, the universal nature of the dynamics is revealed. 
\section{Conclusions}
For the thermalization of a disordered interacting Floquet system, we analyze the dynamics of sub-system temperature $\bej$ {in the ergodic regime}. Generically, some blocks heat up faster than others, but all blocks eventually thermalize, leading to a broad distribution of thermalization times \SB{as identified by the plateau formation in the distribution of decay times with increasing disorder strength}.  
Blocks with long thermalization times are not  particularly \emph{rare} and may exhibit 
either long exponential decay time constants or even nonexponential decay.
This distribution of time scales leads to a slower decay of the spatial and disorder averaged inverse temperature $\bej$, resulting in nonexponential heating over time in the ergodic phase, resembling the relaxation dynamics of classical glassy liquids.
Exploring the connection between inhomogeneous thermalization dynamics and the avalanche mechanism or many-body resonances~\cite{deroeckPRB17, Thiery2017,  GopalakrishnanPRB15, Garratt2021, GarrattPRB22,  Morningstar2022, CrowleySciP22, LongPRL23, LeonardNatPhys23}, which predict stretched exponential decay of correlation functions, is an obvious avenue for further study.

Identifying diagonal entropy $\Sd$ as an \SB{effective} internal system time allows a collapse of the thermalization dynamics across all the disorder values in the ergodic phase, revealing a remarkable, albeit hidden, homogeneity. 
The thermalization process involves shrinking off-diagonal matrix elements of the density matrix. In this sense, the approximation of time with diagonal entropy $\Sd$  measures the heating rate along with the Fock space delocalization.  
\SB{In contrast, for a non-driven model that conserves both particle number and energy, achieving such data collapse requires a fitting parameter, which is demonstrated in App.~\ref{app:nond}.}

The prediction of any dynamical exponent is challenging due to limits imposed by finite time and system sizes, as copiously noted in Hamiltonian models~\cite{Bera2017, Weiner19, PandaMBL19, ChandaPRB20, SierantLargeWc20,  DoggenRevAnnPhy21, EversPRB23}, disordered Floquet models~\cite{TaliaPRB19, LongPRL23}, and even in clean models, where $L\gtrsim 22$ is often necessary to observe heating towards infinite temperature~\cite{RakcheevPRR22}.
Indeed, there is 
substantial variation in stretched exponents with increasing system sizes, particularly evident in two-point correlators, and their fate even in the ergodic phase in the asymptotic limit~\cite{TaliaPRB19, Asmi23} is at this point unclear. We note that the data collapse we observe is largely independent of system size and thus appears less afflicted by finite-size effects.

\section{Acknowledgements}
SB would like to thank F. Evers for several insightful discussions and collaboration on related topics. We would also like to thank C. Artiaco and M. Haque for several discussions. SB acknowledges support from MPG for funding through the Max Planck Partner Group at IIT Bombay and also thanks the MPI-PKS, Dresden computing cluster, where the calculation is performed.  This work was in part supported by the Deutsche Forschungsgemeinschaft  under grant cluster of excellence ct.qmat (EXC 2147, project-id 390858490). IM acknowledges financial support from the Prime Minister’s Research Fellows (PMRF) scheme offered by the Ministry of Education, Government of India.

\appendix

\section{Finite $L$  dependence}
\label{app:fs}
Figure~\ref{fig:betaL} upper panel shows the system size dependence of the decay of the disorder averaged inverse local temperature $\overline{\beta}(n)$. 
With decreasing $\Ga$, i.e., increasing the strength of the disorder, the $L$-dependence is more severe for finite time simulations.
The system thermalizes quickly for larger $\Ga=0.7$, and the data shows less finite-size corrections.

\section{Estimation of decay time $\tau_j$}
\label{app:tau_fit}
\begin{figure}[!tb]
    \centering
    \includegraphics[width=1.0\columnwidth]
    {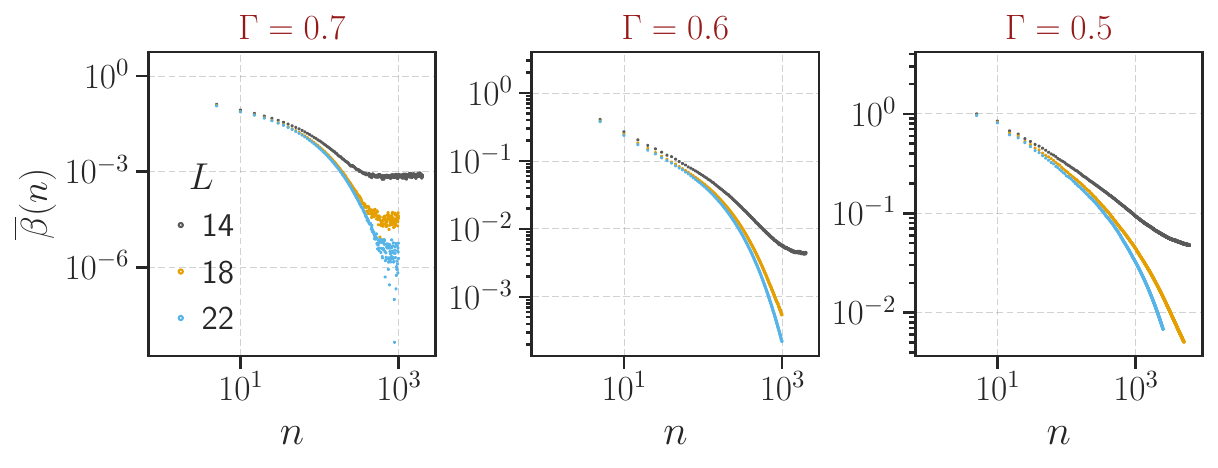}
    \includegraphics[width=1.0\columnwidth]{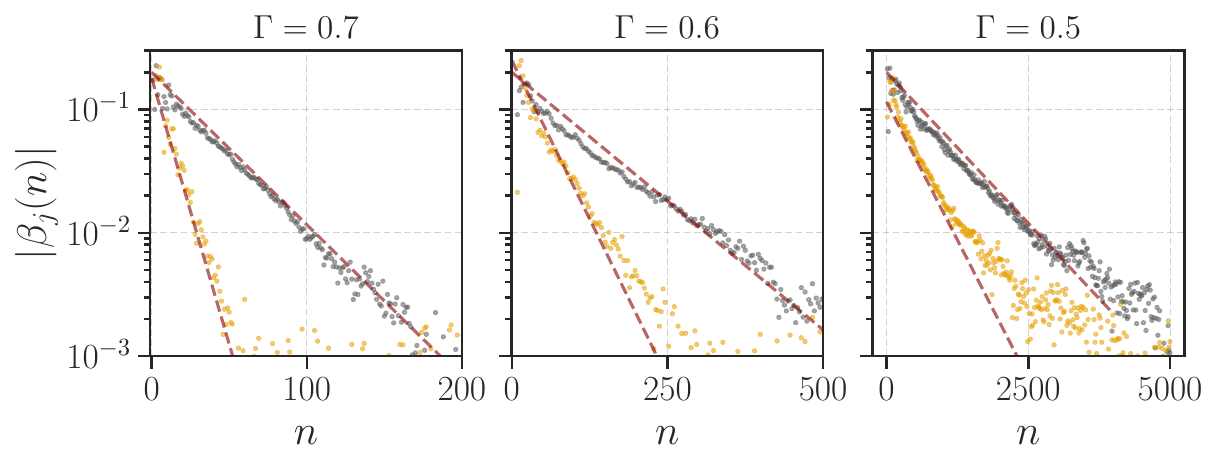}
    \caption{Upper panel: $L$-dependence of the decay of $\beta(n)$ with $n$ for different values of $\Gamma=0.7, 0.6, 0.5$. Lower panel: Decay of  $\bej$ for two typical disorder configurations. The dashed lines show the curve corresponding to the estimated decay times defined in Eq.~\eqref{eq:tau}.}
    \label{fig:betaL}
\end{figure}

Figure~\ref{fig:betaL} lower panel shows the decay of  $\bej$ for two typical disorder configurations for different values of disorder strength, and $L=20$. 
The dashed line represents the estimated curve using the decay time $\tau_j$ defined in Eq.~\eqref{eq:tau}.
The estimate of $\tau_j$ reasonably reproduces the decay for pure exponential traces. 
When $\bej$ shows non-exponential decay for a stronger disorder, the $\tau_j$ gives only a rough estimate as visible at smaller $\Gamma \gtrsim 0.5$.

\begin{figure}[!tb]
    \centering
    \includegraphics[width=1\columnwidth]{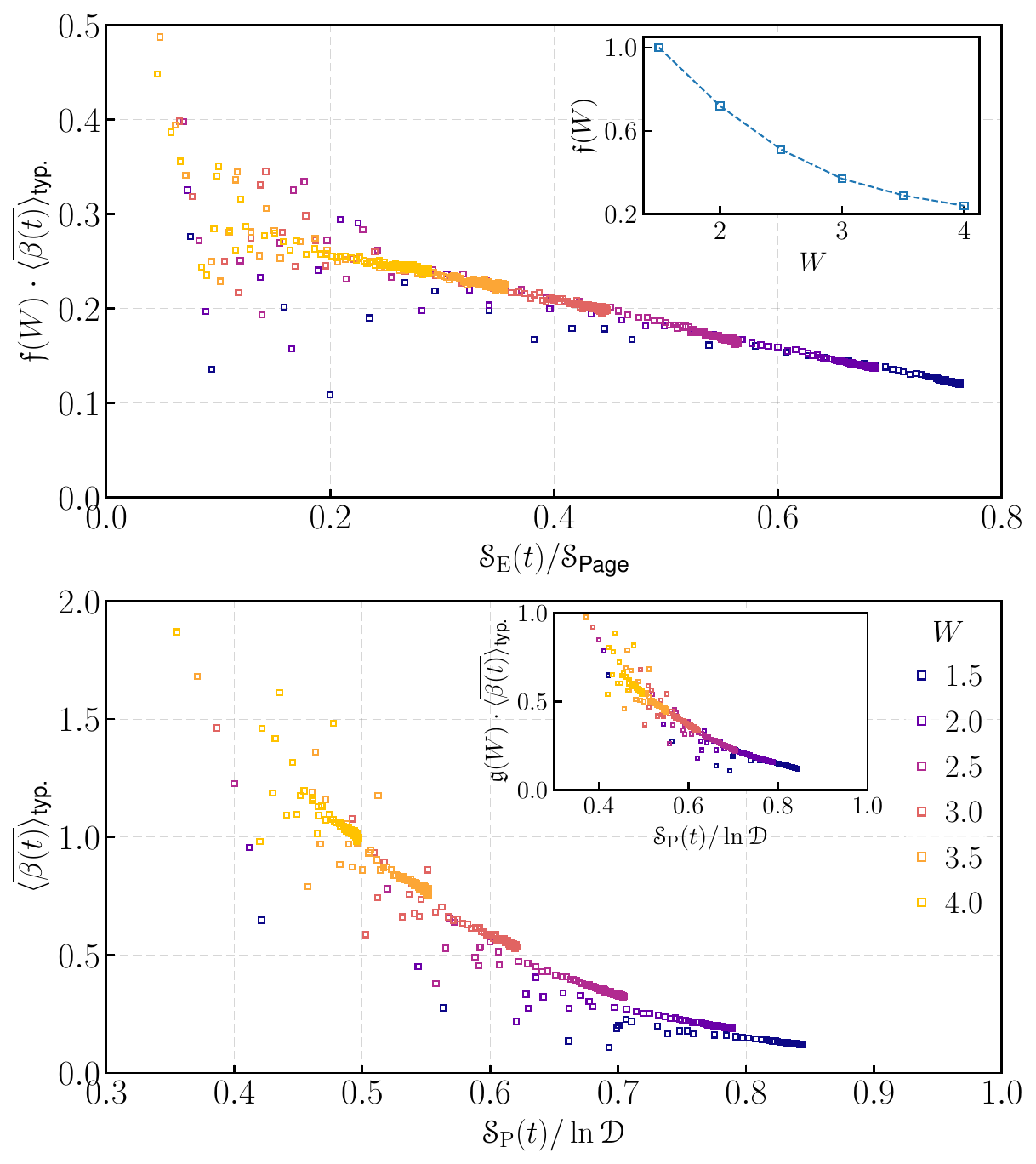}
    \caption{Shows the evolution of the disorder averaged typical inverse local temperature $x(t) = \la \overline{\beta(t)}\ra_\text{typ.}$ for several disorder values $W$ for $L=16$ for the non-driven XXZ model. 
    Upper panel: Shows the $x(t)$ when time is replaced by bipartite entanglement entropy $\msr{S}_\text{E}$ normalized with the Page value. An approximate collapse is achieved with one fitting parameter that depends on the disorder strength, $\mathfrak{f}(W)$. The inset shows the $\mathfrak{f}(W)$ as a function of $W$. 
     Below: The evolution of the typical inverse temperature when participation entropy $\mSP$ normalized with the Fock space dimension $\mathscr{D}$ is adapted as an implicit measure of time, $t$, for the above figure. Inset shows an approximate collapse of the data with one fitting parameter $\mathfrak{g}(W)$.}
    \label{fig:xxz_beta_collapse}
\end{figure}

\section{Proof of $\mSP =2 \Sd$}
\label{app:sd}
Here, we establish the relation between the Fock space~(FS) participation entropy $\mathscr{S}_{\text{P}}$ and the diagonal entropy in the FS basis defined in Eq.~\eqref{eq:sdiag}. 
\begin{align*}
    \msr{S}_{\mathrm{P}} = -\sum_{i=1}^{2^L} |c_i(t)|^2 \log (|c_i(t)|^2) 
\end{align*}
from the definition of the participation entropy. 
Now, the probabilities, $|c_i(t)|^2$ can be written in terms of the diagonal elements of the density matrix of the system, $\rho$ in the following way. We note,  $|c_i(t)|^2 = c_i(t) c_i^*(t) = \la i |\psi\ra \la \psi | i \ra = \la i | \rho | i \ra$,
where $\rho = |\psi \ra \la \psi |$ is the system's density matrix.
\begin{equation*}
    \mathscr{S}_{\mathrm{P}} = - \sum_{i = 1}^{2^L} \rho_{ii} \log \rho_{ii}
\end{equation*}
Here, we can decompose the i-th diagonal element of the full density matrix, i.e., $\rho_{ii}$, as a product of diagonal elements of the sub-system density matrices.
\eq{
\label{decomposition of rho into rhoA and rhoB}
    \rho_{ii} & =   \la i | \rho_A \otimes \rho_B | i \ra  \nn \\
    &= \la j | \la k| \rho_A \otimes \rho_B | j \ra | k \ra \nn \\
    & = \rho_{jj}^A \rho_{kk}^B 
}
here, $|i\ra = |j\ra |k\ra$. 
Therefore, 
\eq{
    \mathscr{S}_{\mathrm{P}} & = - \sum_{i = 1}^{2^L} \rho_{ii} \log \rho_{ii} \nn  \\
     & = \sum_{j=1}^{2^{L/2}} \sum_{k=1}^{2^{L/2}} \rho_{jj}^A \rho_{kk}^B \log (\rho_{jj}^A \rho_{kk}^B) \hspace{18 mm} [\text{using Eq. \ref{decomposition of rho into rhoA and rhoB}}] \nn \\
     & = \sum_{j=1}^{2^{L/2}} \sum_{k=1}^{2^{L/2}} \rho_{jj}^A \rho_{kk}^B \left (\log (\rho_{jj}^A) + \log( \rho_{kk}^B) \right)\nn \\
     &  = \sum_{k=1}^{2^{L/2}} \rho_{kk}^B \sum_{j=1}^{2^{L/2}} \rho_{jj}^A \log (\rho_{jj}^A)  + \sum_{j=1}^{2^{L/2}} \rho_{jj}^A \sum_{k=1}^{2^{L/2}} \rho_{kk}^B \log (\rho_{kk}^B) \nn \\
     & = 2 \sum_{j=1}^{2^{L/2}} \rho_{jj}^{A} \log (\rho_{jj}^A) \nn \\
     & = 2 \mathscr{S}_{\mathrm{d}}
}

\SB{
\section{Non-driven model}
\label{app:nond}
This section considers a model with particle number and energy conservation, namely the XXZ model with the random magnetic field and with open boundary condition, 
\eq{
\msr{H} = \sum_j^{L-1} {\bf S}_j \cdot {\bf S}_{j+1} + \sum_j^{L} h_j  S_j^z. 
}
The $h_j$ is the random field from a uniform box distribution $[-W, W]$. We consider a quench from the Neel state and observe the entanglement growth and the local temperature $\bej$. \newline
The upper panel of Fig.Fig.~\ref{fig:xxz_beta_collapse} shows the evolution of the typical inverse local temperature, $x(t) = \la \overline{\beta(t)}\ra_\text{typ.}$, as a function of the entanglement entropy $\mSE$. Using a single fitting parameter that depends on the disorder strength $W$, we observe an approximate data collapse, indicating the homogeneity of the ergodic phase. The inset illustrates the dependence of the fitting parameter $\mathfrak{f}(W)$ on the disorder, showing a decay with increasing disorder. \newline
The lower panel presents the same data, $x(t)$, plotted against the Fock space participation entropy $\mSP$. In contrast to the driven Ising model, we do not observe a data collapse in this case, which may be attributed to the conservation of particle number and energy. However, as shown in the inset, allowing for a single fitting parameter results in an approximate collapse of the time evolution, indicating the homogeneous dynamics in the evolution of the typical inverse temperature, which might otherwise remain obscured.
}

\bibliography{MBL}

\begin{thebibliography}{60}%
\makeatletter
\providecommand \@ifxundefined [1]{%
 \@ifx{#1\undefined}
}%
\providecommand \@ifnum [1]{%
 \ifnum #1\expandafter \@firstoftwo
 \else \expandafter \@secondoftwo
 \fi
}%
\providecommand \@ifx [1]{%
 \ifx #1\expandafter \@firstoftwo
 \else \expandafter \@secondoftwo
 \fi
}%
\providecommand \natexlab [1]{#1}%
\providecommand \enquote  [1]{``#1''}%
\providecommand \bibnamefont  [1]{#1}%
\providecommand \bibfnamefont [1]{#1}%
\providecommand \citenamefont [1]{#1}%
\providecommand \href@noop [0]{\@secondoftwo}%
\providecommand \href [0]{\begingroup \@sanitize@url \@href}%
\providecommand \@href[1]{\@@startlink{#1}\@@href}%
\providecommand \@@href[1]{\endgroup#1\@@endlink}%
\providecommand \@sanitize@url [0]{\catcode `\\12\catcode `\$12\catcode `\&12\catcode `\#12\catcode `\^12\catcode `\_12\catcode `\%12\relax}%
\providecommand \@@startlink[1]{}%
\providecommand \@@endlink[0]{}%
\providecommand \url  [0]{\begingroup\@sanitize@url \@url }%
\providecommand \@url [1]{\endgroup\@href {#1}{\urlprefix }}%
\providecommand \urlprefix  [0]{URL }%
\providecommand \Eprint [0]{\href }%
\providecommand \doibase [0]{https://doi.org/}%
\providecommand \selectlanguage [0]{\@gobble}%
\providecommand \bibinfo  [0]{\@secondoftwo}%
\providecommand \bibfield  [0]{\@secondoftwo}%
\providecommand \translation [1]{[#1]}%
\providecommand \BibitemOpen [0]{}%
\providecommand \bibitemStop [0]{}%
\providecommand \bibitemNoStop [0]{.\EOS\space}%
\providecommand \EOS [0]{\spacefactor3000\relax}%
\providecommand \BibitemShut  [1]{\csname bibitem#1\endcsname}%
\let\auto@bib@innerbib\@empty
\bibitem [{\citenamefont {Berthier}\ and\ \citenamefont {Biroli}(2011)}]{BerthierRMP11}%
  \BibitemOpen
  \bibfield  {author} {\bibinfo {author} {\bibfnamefont {L.}~\bibnamefont {Berthier}}\ and\ \bibinfo {author} {\bibfnamefont {G.}~\bibnamefont {Biroli}},\ }\bibfield  {title} {\bibinfo {title} {Theoretical perspective on the glass transition and amorphous materials},\ }\href {https://doi.org/10.1103/RevModPhys.83.587} {\bibfield  {journal} {\bibinfo  {journal} {Rev. Mod. Phys.}\ }\textbf {\bibinfo {volume} {83}},\ \bibinfo {pages} {587} (\bibinfo {year} {2011})}\BibitemShut {NoStop}%
\bibitem [{\citenamefont {Glotzer}\ \emph {et~al.}(1998)\citenamefont {Glotzer}, \citenamefont {Jan}, \citenamefont {Lookman}, \citenamefont {MacIsaac},\ and\ \citenamefont {Poole}}]{GlotzerPRE98}%
  \BibitemOpen
  \bibfield  {author} {\bibinfo {author} {\bibfnamefont {S.~C.}\ \bibnamefont {Glotzer}}, \bibinfo {author} {\bibfnamefont {N.}~\bibnamefont {Jan}}, \bibinfo {author} {\bibfnamefont {T.}~\bibnamefont {Lookman}}, \bibinfo {author} {\bibfnamefont {A.~B.}\ \bibnamefont {MacIsaac}},\ and\ \bibinfo {author} {\bibfnamefont {P.~H.}\ \bibnamefont {Poole}},\ }\bibfield  {title} {\bibinfo {title} {Dynamical heterogeneity in the ising spin glass},\ }\href {https://doi.org/10.1103/PhysRevE.57.7350} {\bibfield  {journal} {\bibinfo  {journal} {Phys. Rev. E}\ }\textbf {\bibinfo {volume} {57}},\ \bibinfo {pages} {7350} (\bibinfo {year} {1998})}\BibitemShut {NoStop}%
\bibitem [{\citenamefont {Ediger}(2000)}]{EdigerARPC00}%
  \BibitemOpen
  \bibfield  {author} {\bibinfo {author} {\bibfnamefont {M.~D.}\ \bibnamefont {Ediger}},\ }\bibfield  {title} {\bibinfo {title} {Spatially heterogeneous dynamics in supercooled liquids},\ }\href {https://doi.org/10.1146/annurev.physchem.51.1.99} {\bibfield  {journal} {\bibinfo  {journal} {Annu. Rev. Phys. Chem.}\ }\textbf {\bibinfo {volume} {51}},\ \bibinfo {pages} {99} (\bibinfo {year} {2000})}\BibitemShut {NoStop}%
\bibitem [{\citenamefont {Nandkishore}\ and\ \citenamefont {Huse}(2015)}]{Nandkishore2015}%
  \BibitemOpen
  \bibfield  {author} {\bibinfo {author} {\bibfnamefont {R.}~\bibnamefont {Nandkishore}}\ and\ \bibinfo {author} {\bibfnamefont {D.~A.}\ \bibnamefont {Huse}},\ }\bibfield  {title} {\bibinfo {title} {Many-body localization and thermalization in quantum statistical mechanics},\ }\href {https://doi.org/10.1146/annurev-conmatphys-031214-014726} {\bibfield  {journal} {\bibinfo  {journal} {Ann. Rev. Condens. Matter Phys.}\ }\textbf {\bibinfo {volume} {6}},\ \bibinfo {pages} {15} (\bibinfo {year} {2015})}\BibitemShut {NoStop}%
\bibitem [{\citenamefont {D'Alessio}\ \emph {et~al.}(2016)\citenamefont {D'Alessio}, \citenamefont {Kafri}, \citenamefont {Polkovnikov},\ and\ \citenamefont {Rigol}}]{DAlessio2016}%
  \BibitemOpen
  \bibfield  {author} {\bibinfo {author} {\bibfnamefont {L.}~\bibnamefont {D'Alessio}}, \bibinfo {author} {\bibfnamefont {Y.}~\bibnamefont {Kafri}}, \bibinfo {author} {\bibfnamefont {A.}~\bibnamefont {Polkovnikov}},\ and\ \bibinfo {author} {\bibfnamefont {M.}~\bibnamefont {Rigol}},\ }\bibfield  {title} {\bibinfo {title} {From quantum chaos and eigenstate thermalization to statistical mechanics and thermodynamics},\ }\href {https://doi.org/10.1080/00018732.2016.1198134} {\bibfield  {journal} {\bibinfo  {journal} {Adv. Phys.}\ }\textbf {\bibinfo {volume} {65}},\ \bibinfo {pages} {239} (\bibinfo {year} {2016})}\BibitemShut {NoStop}%
\bibitem [{\citenamefont {Ueda}(2020)}]{UedaNat2020}%
  \BibitemOpen
  \bibfield  {author} {\bibinfo {author} {\bibfnamefont {M.}~\bibnamefont {Ueda}},\ }\bibfield  {title} {\bibinfo {title} {Quantum equilibration, thermalization and prethermalization in ultracold atoms},\ }\href {https://doi.org/10.1038/s42254-020-0237-x} {\bibfield  {journal} {\bibinfo  {journal} {Nature Reviews Physics}\ }\textbf {\bibinfo {volume} {2}},\ \bibinfo {pages} {669} (\bibinfo {year} {2020})}\BibitemShut {NoStop}%
\bibitem [{\citenamefont {Gornyi}\ \emph {et~al.}(2005)\citenamefont {Gornyi}, \citenamefont {Mirlin},\ and\ \citenamefont {Polyakov}}]{Gornyi2005}%
  \BibitemOpen
  \bibfield  {author} {\bibinfo {author} {\bibfnamefont {I.~V.}\ \bibnamefont {Gornyi}}, \bibinfo {author} {\bibfnamefont {A.~D.}\ \bibnamefont {Mirlin}},\ and\ \bibinfo {author} {\bibfnamefont {D.~G.}\ \bibnamefont {Polyakov}},\ }\bibfield  {title} {\bibinfo {title} {Interacting electrons in disordered wires: Anderson localization and low-$t$ transport},\ }\href {https://doi.org/10.1103/PhysRevLett.95.206603} {\bibfield  {journal} {\bibinfo  {journal} {Phys. Rev. Lett.}\ }\textbf {\bibinfo {volume} {95}},\ \bibinfo {pages} {206603} (\bibinfo {year} {2005})}\BibitemShut {NoStop}%
\bibitem [{\citenamefont {Basko}\ \emph {et~al.}(2006)\citenamefont {Basko}, \citenamefont {Aleiner},\ and\ \citenamefont {Altshuler}}]{Basko2006}%
  \BibitemOpen
  \bibfield  {author} {\bibinfo {author} {\bibfnamefont {D.~M.}\ \bibnamefont {Basko}}, \bibinfo {author} {\bibfnamefont {I.~L.}\ \bibnamefont {Aleiner}},\ and\ \bibinfo {author} {\bibfnamefont {B.~L.}\ \bibnamefont {Altshuler}},\ }\bibfield  {title} {\bibinfo {title} {Metal insulator transition in a weakly interacting many electron system with localized single particle states},\ }\href {http://www.sciencedirect.com/science/article/pii/S0003491605002630} {\bibfield  {journal} {\bibinfo  {journal} {Ann. Phys.}\ }\textbf {\bibinfo {volume} {321}},\ \bibinfo {pages} {1126 } (\bibinfo {year} {2006})}\BibitemShut {NoStop}%
\bibitem [{\citenamefont {Bera}\ \emph {et~al.}(2015)\citenamefont {Bera}, \citenamefont {Schomerus}, \citenamefont {Heidrich-Meisner},\ and\ \citenamefont {Bardarson}}]{Bera2015}%
  \BibitemOpen
  \bibfield  {author} {\bibinfo {author} {\bibfnamefont {S.}~\bibnamefont {Bera}}, \bibinfo {author} {\bibfnamefont {H.}~\bibnamefont {Schomerus}}, \bibinfo {author} {\bibfnamefont {F.}~\bibnamefont {Heidrich-Meisner}},\ and\ \bibinfo {author} {\bibfnamefont {J.~H.}\ \bibnamefont {Bardarson}},\ }\bibfield  {title} {\bibinfo {title} {Many-body localization characterized from a one-particle perspective},\ }\href {https://doi.org/10.1103/PhysRevLett.115.046603} {\bibfield  {journal} {\bibinfo  {journal} {Phys. Rev. Lett.}\ }\textbf {\bibinfo {volume} {115}},\ \bibinfo {pages} {046603} (\bibinfo {year} {2015})}\BibitemShut {NoStop}%
\bibitem [{\citenamefont {Abanin}\ \emph {et~al.}(2019)\citenamefont {Abanin}, \citenamefont {Altman}, \citenamefont {Bloch},\ and\ \citenamefont {Serbyn}}]{AbaninBloch-Review-2018}%
  \BibitemOpen
  \bibfield  {author} {\bibinfo {author} {\bibfnamefont {D.~A.}\ \bibnamefont {Abanin}}, \bibinfo {author} {\bibfnamefont {E.}~\bibnamefont {Altman}}, \bibinfo {author} {\bibfnamefont {I.}~\bibnamefont {Bloch}},\ and\ \bibinfo {author} {\bibfnamefont {M.}~\bibnamefont {Serbyn}},\ }\bibfield  {title} {\bibinfo {title} {Colloquium: Many-body localization, thermalization, and entanglement},\ }\href {https://doi.org/10.1103/RevModPhys.91.021001} {\bibfield  {journal} {\bibinfo  {journal} {Rev. Mod. Phys.}\ }\textbf {\bibinfo {volume} {91}},\ \bibinfo {pages} {021001} (\bibinfo {year} {2019})}\BibitemShut {NoStop}%
\bibitem [{\citenamefont {Alet}\ and\ \citenamefont {Laflorencie}(2018)}]{AletReview2018}%
  \BibitemOpen
  \bibfield  {author} {\bibinfo {author} {\bibfnamefont {F.}~\bibnamefont {Alet}}\ and\ \bibinfo {author} {\bibfnamefont {N.}~\bibnamefont {Laflorencie}},\ }\bibfield  {title} {\bibinfo {title} {Many-body localization: An introduction and selected topics},\ }\href {http://www.sciencedirect.com/science/article/pii/S163107051830032X} {\bibfield  {journal} {\bibinfo  {journal} {C. R. Phys.}\ } (\bibinfo {year} {2018})}\BibitemShut {NoStop}%
\bibitem [{\citenamefont {Sierant}\ \emph {et~al.}(2025)\citenamefont {Sierant}, \citenamefont {Lewenstein}, \citenamefont {Scardicchio}, \citenamefont {Vidmar},\ and\ \citenamefont {Zakrzewski}}]{Sierant_2025}%
  \BibitemOpen
  \bibfield  {author} {\bibinfo {author} {\bibfnamefont {P.}~\bibnamefont {Sierant}}, \bibinfo {author} {\bibfnamefont {M.}~\bibnamefont {Lewenstein}}, \bibinfo {author} {\bibfnamefont {A.}~\bibnamefont {Scardicchio}}, \bibinfo {author} {\bibfnamefont {L.}~\bibnamefont {Vidmar}},\ and\ \bibinfo {author} {\bibfnamefont {J.}~\bibnamefont {Zakrzewski}},\ }\bibfield  {title} {\bibinfo {title} {Many-body localization in the age of classical computing},\ }\href {https://doi.org/10.1088/1361-6633/ad9756} {\bibfield  {journal} {\bibinfo  {journal} {Reports on Progress in Physics}\ }\textbf {\bibinfo {volume} {88}},\ \bibinfo {pages} {026502} (\bibinfo {year} {2025})}\BibitemShut {NoStop}%
\bibitem [{\citenamefont {Bera}\ and\ \citenamefont {Lakshminarayan}(2016)}]{BeraArul16}%
  \BibitemOpen
  \bibfield  {author} {\bibinfo {author} {\bibfnamefont {S.}~\bibnamefont {Bera}}\ and\ \bibinfo {author} {\bibfnamefont {A.}~\bibnamefont {Lakshminarayan}},\ }\bibfield  {title} {\bibinfo {title} {Local entanglement structure across a many-body localization transition},\ }\href {https://doi.org/10.1103/PhysRevB.93.134204} {\bibfield  {journal} {\bibinfo  {journal} {Phys. Rev. B}\ }\textbf {\bibinfo {volume} {93}},\ \bibinfo {pages} {134204} (\bibinfo {year} {2016})}\BibitemShut {NoStop}%
\bibitem [{\citenamefont {Artiaco}\ \emph {et~al.}(2022)\citenamefont {Artiaco}, \citenamefont {Balducci}, \citenamefont {Heyl}, \citenamefont {Russomanno},\ and\ \citenamefont {Scardicchio}}]{ClaudiaPRB22}%
  \BibitemOpen
  \bibfield  {author} {\bibinfo {author} {\bibfnamefont {C.}~\bibnamefont {Artiaco}}, \bibinfo {author} {\bibfnamefont {F.}~\bibnamefont {Balducci}}, \bibinfo {author} {\bibfnamefont {M.}~\bibnamefont {Heyl}}, \bibinfo {author} {\bibfnamefont {A.}~\bibnamefont {Russomanno}},\ and\ \bibinfo {author} {\bibfnamefont {A.}~\bibnamefont {Scardicchio}},\ }\bibfield  {title} {\bibinfo {title} {Spatiotemporal heterogeneity of entanglement in many-body localized systems},\ }\href {https://doi.org/10.1103/PhysRevB.105.184202} {\bibfield  {journal} {\bibinfo  {journal} {Phys. Rev. B}\ }\textbf {\bibinfo {volume} {105}},\ \bibinfo {pages} {184202} (\bibinfo {year} {2022})}\BibitemShut {NoStop}%
\bibitem [{\citenamefont {Khemani}\ \emph {et~al.}(2017)\citenamefont {Khemani}, \citenamefont {Lim}, \citenamefont {Sheng},\ and\ \citenamefont {Huse}}]{Khemani2017}%
  \BibitemOpen
  \bibfield  {author} {\bibinfo {author} {\bibfnamefont {V.}~\bibnamefont {Khemani}}, \bibinfo {author} {\bibfnamefont {S.~P.}\ \bibnamefont {Lim}}, \bibinfo {author} {\bibfnamefont {D.~N.}\ \bibnamefont {Sheng}},\ and\ \bibinfo {author} {\bibfnamefont {D.~A.}\ \bibnamefont {Huse}},\ }\bibfield  {title} {\bibinfo {title} {Critical properties of the many-body localization transition},\ }\href {https://doi.org/10.1103/PhysRevX.7.021013} {\bibfield  {journal} {\bibinfo  {journal} {Phys. Rev. X}\ }\textbf {\bibinfo {volume} {7}},\ \bibinfo {pages} {021013} (\bibinfo {year} {2017})}\BibitemShut {NoStop}%
\bibitem [{\citenamefont {Weiner}\ \emph {et~al.}(2019)\citenamefont {Weiner}, \citenamefont {Evers},\ and\ \citenamefont {Bera}}]{Weiner19}%
  \BibitemOpen
  \bibfield  {author} {\bibinfo {author} {\bibfnamefont {F.}~\bibnamefont {Weiner}}, \bibinfo {author} {\bibfnamefont {F.}~\bibnamefont {Evers}},\ and\ \bibinfo {author} {\bibfnamefont {S.}~\bibnamefont {Bera}},\ }\bibfield  {title} {\bibinfo {title} {Slow dynamics and strong finite-size effects in many-body localization with random and quasiperiodic potentials},\ }\href {https://doi.org/10.1103/PhysRevB.100.104204} {\bibfield  {journal} {\bibinfo  {journal} {Phys. Rev. B}\ }\textbf {\bibinfo {volume} {100}},\ \bibinfo {pages} {104204} (\bibinfo {year} {2019})}\BibitemShut {NoStop}%
\bibitem [{\citenamefont {Panda}\ \emph {et~al.}(2020)\citenamefont {Panda}, \citenamefont {Scardicchio}, \citenamefont {Schulz}, \citenamefont {Taylor},\ and\ \citenamefont {{\v{Z}}nidari{\v{c}}}}]{PandaMBL19}%
  \BibitemOpen
  \bibfield  {author} {\bibinfo {author} {\bibfnamefont {R.~K.}\ \bibnamefont {Panda}}, \bibinfo {author} {\bibfnamefont {A.}~\bibnamefont {Scardicchio}}, \bibinfo {author} {\bibfnamefont {M.}~\bibnamefont {Schulz}}, \bibinfo {author} {\bibfnamefont {S.~R.}\ \bibnamefont {Taylor}},\ and\ \bibinfo {author} {\bibfnamefont {M.}~\bibnamefont {{\v{Z}}nidari{\v{c}}}},\ }\bibfield  {title} {\bibinfo {title} {Can we study the many-body localisation transition?},\ }\href {https://doi.org/10.1209/0295-5075/128/67003} {\bibfield  {journal} {\bibinfo  {journal} {{EPL} (Europhysics Letters)}\ }\textbf {\bibinfo {volume} {128}},\ \bibinfo {pages} {67003} (\bibinfo {year} {2020})}\BibitemShut {NoStop}%
\bibitem [{\citenamefont {Kiefer-Emmanouilidis}\ \emph {et~al.}(2020)\citenamefont {Kiefer-Emmanouilidis}, \citenamefont {Unanyan}, \citenamefont {Fleischhauer},\ and\ \citenamefont {Sirker}}]{SirkerPRL20}%
  \BibitemOpen
  \bibfield  {author} {\bibinfo {author} {\bibfnamefont {M.}~\bibnamefont {Kiefer-Emmanouilidis}}, \bibinfo {author} {\bibfnamefont {R.}~\bibnamefont {Unanyan}}, \bibinfo {author} {\bibfnamefont {M.}~\bibnamefont {Fleischhauer}},\ and\ \bibinfo {author} {\bibfnamefont {J.}~\bibnamefont {Sirker}},\ }\bibfield  {title} {\bibinfo {title} {Evidence for unbounded growth of the number entropy in many-body localized phases},\ }\href {https://doi.org/10.1103/PhysRevLett.124.243601} {\bibfield  {journal} {\bibinfo  {journal} {Phys. Rev. Lett.}\ }\textbf {\bibinfo {volume} {124}},\ \bibinfo {pages} {243601} (\bibinfo {year} {2020})}\BibitemShut {NoStop}%
\bibitem [{\citenamefont {Luitz}\ and\ \citenamefont {Lev}(2020)}]{Luitz2020}%
  \BibitemOpen
  \bibfield  {author} {\bibinfo {author} {\bibfnamefont {D.~J.}\ \bibnamefont {Luitz}}\ and\ \bibinfo {author} {\bibfnamefont {Y.~B.}\ \bibnamefont {Lev}},\ }\bibfield  {title} {\bibinfo {title} {{Absence of slow particle transport in the many-body localized phase}},\ }\href {https://doi.org/10.1103/PhysRevB.102.100202} {\bibfield  {journal} {\bibinfo  {journal} {Phys. Rev. B}\ }\textbf {\bibinfo {volume} {102}},\ \bibinfo {pages} {100202} (\bibinfo {year} {2020})}\BibitemShut {NoStop}%
\bibitem [{\citenamefont {\v{S}untajs}\ \emph {et~al.}(2020)\citenamefont {\v{S}untajs}, \citenamefont {Bon\v{c}a}, \citenamefont {Prosen},\ and\ \citenamefont {Vidmar}}]{SuntasPRB20}%
  \BibitemOpen
  \bibfield  {author} {\bibinfo {author} {\bibfnamefont {J.}~\bibnamefont {\v{S}untajs}}, \bibinfo {author} {\bibfnamefont {J.}~\bibnamefont {Bon\v{c}a}}, \bibinfo {author} {\bibfnamefont {T.}~\bibnamefont {Prosen}},\ and\ \bibinfo {author} {\bibfnamefont {L.}~\bibnamefont {Vidmar}},\ }\bibfield  {title} {\bibinfo {title} {Ergodicity breaking transition in finite disordered spin chains},\ }\href {https://doi.org/10.1103/PhysRevB.102.064207} {\bibfield  {journal} {\bibinfo  {journal} {Phys. Rev. B}\ }\textbf {\bibinfo {volume} {102}},\ \bibinfo {pages} {064207} (\bibinfo {year} {2020})}\BibitemShut {NoStop}%
\bibitem [{\citenamefont {Abanin}\ \emph {et~al.}(2021)\citenamefont {Abanin}, \citenamefont {Bardarson}, \citenamefont {{De Tomasi}}, \citenamefont {Gopalakrishnan}, \citenamefont {Khemani}, \citenamefont {Parameswaran}, \citenamefont {Pollmann}, \citenamefont {Potter}, \citenamefont {Serbyn},\ and\ \citenamefont {Vasseur}}]{AbaninAOP21}%
  \BibitemOpen
  \bibfield  {author} {\bibinfo {author} {\bibfnamefont {D.}~\bibnamefont {Abanin}}, \bibinfo {author} {\bibfnamefont {J.}~\bibnamefont {Bardarson}}, \bibinfo {author} {\bibfnamefont {G.}~\bibnamefont {{De Tomasi}}}, \bibinfo {author} {\bibfnamefont {S.}~\bibnamefont {Gopalakrishnan}}, \bibinfo {author} {\bibfnamefont {V.}~\bibnamefont {Khemani}}, \bibinfo {author} {\bibfnamefont {S.}~\bibnamefont {Parameswaran}}, \bibinfo {author} {\bibfnamefont {F.}~\bibnamefont {Pollmann}}, \bibinfo {author} {\bibfnamefont {A.}~\bibnamefont {Potter}}, \bibinfo {author} {\bibfnamefont {M.}~\bibnamefont {Serbyn}},\ and\ \bibinfo {author} {\bibfnamefont {R.}~\bibnamefont {Vasseur}},\ }\bibfield  {title} {\bibinfo {title} {{Distinguishing localization from chaos: Challenges in finite-size systems}},\ }\href {https://doi.org/https://doi.org/10.1016/j.aop.2021.168415} {\bibfield  {journal} {\bibinfo  {journal} {Annals of Physics}\ }\textbf {\bibinfo {volume} {427}},\ \bibinfo {pages} {168415} (\bibinfo {year}
  {2021})}\BibitemShut {NoStop}%
\bibitem [{\citenamefont {Sels}\ and\ \citenamefont {Polkovnikov}(2021)}]{Polkovnikov2021}%
  \BibitemOpen
  \bibfield  {author} {\bibinfo {author} {\bibfnamefont {D.}~\bibnamefont {Sels}}\ and\ \bibinfo {author} {\bibfnamefont {A.}~\bibnamefont {Polkovnikov}},\ }\bibfield  {title} {\bibinfo {title} {Dynamical obstruction to localization in a disordered spin chain},\ }\href {https://doi.org/https://doi.org/10.1103/PhysRevE.104.054105} {\bibfield  {journal} {\bibinfo  {journal} {Phys. Rev. E}\ }\textbf {\bibinfo {volume} {104}},\ \bibinfo {pages} {054105} (\bibinfo {year} {2021})}\BibitemShut {NoStop}%
\bibitem [{\citenamefont {Sels}(2022)}]{SelsBathPRB22}%
  \BibitemOpen
  \bibfield  {author} {\bibinfo {author} {\bibfnamefont {D.}~\bibnamefont {Sels}},\ }\bibfield  {title} {\bibinfo {title} {Bath-induced delocalization in interacting disordered spin chains},\ }\href {https://doi.org/10.1103/PhysRevB.106.L020202} {\bibfield  {journal} {\bibinfo  {journal} {Phys. Rev. B}\ }\textbf {\bibinfo {volume} {106}},\ \bibinfo {pages} {L020202} (\bibinfo {year} {2022})}\BibitemShut {NoStop}%
\bibitem [{\citenamefont {Sierant}\ and\ \citenamefont {Zakrzewski}(2022)}]{SierantPRB22}%
  \BibitemOpen
  \bibfield  {author} {\bibinfo {author} {\bibfnamefont {P.}~\bibnamefont {Sierant}}\ and\ \bibinfo {author} {\bibfnamefont {J.}~\bibnamefont {Zakrzewski}},\ }\bibfield  {title} {\bibinfo {title} {Challenges to observation of many-body localization},\ }\href {https://doi.org/10.1103/PhysRevB.105.224203} {\bibfield  {journal} {\bibinfo  {journal} {Phys. Rev. B}\ }\textbf {\bibinfo {volume} {105}},\ \bibinfo {pages} {224203} (\bibinfo {year} {2022})}\BibitemShut {NoStop}%
\bibitem [{\citenamefont {Morningstar}\ \emph {et~al.}(2022)\citenamefont {Morningstar}, \citenamefont {Colmenarez}, \citenamefont {Khemani}, \citenamefont {Luitz},\ and\ \citenamefont {Huse}}]{Morningstar2022}%
  \BibitemOpen
  \bibfield  {author} {\bibinfo {author} {\bibfnamefont {A.}~\bibnamefont {Morningstar}}, \bibinfo {author} {\bibfnamefont {L.}~\bibnamefont {Colmenarez}}, \bibinfo {author} {\bibfnamefont {V.}~\bibnamefont {Khemani}}, \bibinfo {author} {\bibfnamefont {D.~J.}\ \bibnamefont {Luitz}},\ and\ \bibinfo {author} {\bibfnamefont {D.~A.}\ \bibnamefont {Huse}},\ }\bibfield  {title} {\bibinfo {title} {Avalanches and many-body resonances in many-body localized systems},\ }\href {https://doi.org/10.1103/PhysRevB.105.174205} {\bibfield  {journal} {\bibinfo  {journal} {Phys. Rev. B}\ }\textbf {\bibinfo {volume} {105}},\ \bibinfo {pages} {174205} (\bibinfo {year} {2022})}\BibitemShut {NoStop}%
\bibitem [{\citenamefont {Crowley}\ and\ \citenamefont {Chandran}(2022)}]{CrowleySciP22}%
  \BibitemOpen
  \bibfield  {author} {\bibinfo {author} {\bibfnamefont {P.~J.~D.}\ \bibnamefont {Crowley}}\ and\ \bibinfo {author} {\bibfnamefont {A.}~\bibnamefont {Chandran}},\ }\bibfield  {title} {\bibinfo {title} {{A constructive theory of the numerically accessible many-body localized to thermal crossover}},\ }\href {https://doi.org/10.21468/SciPostPhys.12.6.201} {\bibfield  {journal} {\bibinfo  {journal} {SciPost Phys.}\ }\textbf {\bibinfo {volume} {12}},\ \bibinfo {pages} {201} (\bibinfo {year} {2022})}\BibitemShut {NoStop}%
\bibitem [{\citenamefont {Long}\ \emph {et~al.}(2023)\citenamefont {Long}, \citenamefont {Crowley}, \citenamefont {Khemani},\ and\ \citenamefont {Chandran}}]{LongPRL23}%
  \BibitemOpen
  \bibfield  {author} {\bibinfo {author} {\bibfnamefont {D.~M.}\ \bibnamefont {Long}}, \bibinfo {author} {\bibfnamefont {P.~J.~D.}\ \bibnamefont {Crowley}}, \bibinfo {author} {\bibfnamefont {V.}~\bibnamefont {Khemani}},\ and\ \bibinfo {author} {\bibfnamefont {A.}~\bibnamefont {Chandran}},\ }\bibfield  {title} {\bibinfo {title} {Phenomenology of the prethermal many-body localized regime},\ }\href {https://doi.org/10.1103/PhysRevLett.131.106301} {\bibfield  {journal} {\bibinfo  {journal} {Phys. Rev. Lett.}\ }\textbf {\bibinfo {volume} {131}},\ \bibinfo {pages} {106301} (\bibinfo {year} {2023})}\BibitemShut {NoStop}%
\bibitem [{\citenamefont {Evers}\ \emph {et~al.}(2023)\citenamefont {Evers}, \citenamefont {Modak},\ and\ \citenamefont {Bera}}]{EversPRB23}%
  \BibitemOpen
  \bibfield  {author} {\bibinfo {author} {\bibfnamefont {F.}~\bibnamefont {Evers}}, \bibinfo {author} {\bibfnamefont {I.}~\bibnamefont {Modak}},\ and\ \bibinfo {author} {\bibfnamefont {S.}~\bibnamefont {Bera}},\ }\bibfield  {title} {\bibinfo {title} {Internal clock of many-body delocalization},\ }\href {https://doi.org/10.1103/PhysRevB.108.134204} {\bibfield  {journal} {\bibinfo  {journal} {Phys. Rev. B}\ }\textbf {\bibinfo {volume} {108}},\ \bibinfo {pages} {134204} (\bibinfo {year} {2023})}\BibitemShut {NoStop}%
\bibitem [{\citenamefont {Chávez}\ \emph {et~al.}(2023)\citenamefont {Chávez}, \citenamefont {Artiaco}, \citenamefont {Kvorning}, \citenamefont {Herviou},\ and\ \citenamefont {Bardarson}}]{ChavezUltraSlow23}%
  \BibitemOpen
  \bibfield  {author} {\bibinfo {author} {\bibfnamefont {D.~A.}\ \bibnamefont {Chávez}}, \bibinfo {author} {\bibfnamefont {C.}~\bibnamefont {Artiaco}}, \bibinfo {author} {\bibfnamefont {T.~K.}\ \bibnamefont {Kvorning}}, \bibinfo {author} {\bibfnamefont {L.}~\bibnamefont {Herviou}},\ and\ \bibinfo {author} {\bibfnamefont {J.~H.}\ \bibnamefont {Bardarson}},\ }\href@noop {} {\bibinfo {title} {Ultraslow growth of number entropy in an l-bit model of many-body localization}} (\bibinfo {year} {2023}),\ \Eprint {https://arxiv.org/abs/2312.13420} {arXiv:2312.13420 [cond-mat.dis-nn]} \BibitemShut {NoStop}%
\bibitem [{\citenamefont {Zhang}\ \emph {et~al.}(2016)\citenamefont {Zhang}, \citenamefont {Khemani},\ and\ \citenamefont {Huse}}]{ZhangFloquetPRB16}%
  \BibitemOpen
  \bibfield  {author} {\bibinfo {author} {\bibfnamefont {L.}~\bibnamefont {Zhang}}, \bibinfo {author} {\bibfnamefont {V.}~\bibnamefont {Khemani}},\ and\ \bibinfo {author} {\bibfnamefont {D.~A.}\ \bibnamefont {Huse}},\ }\bibfield  {title} {\bibinfo {title} {A floquet model for the many-body localization transition},\ }\href {https://doi.org/10.1103/PhysRevB.94.224202} {\bibfield  {journal} {\bibinfo  {journal} {Phys. Rev. B}\ }\textbf {\bibinfo {volume} {94}},\ \bibinfo {pages} {224202} (\bibinfo {year} {2016})}\BibitemShut {NoStop}%
\bibitem [{\citenamefont {Lezama}\ \emph {et~al.}(2019)\citenamefont {Lezama}, \citenamefont {Bera},\ and\ \citenamefont {Bardarson}}]{TaliaPRB19}%
  \BibitemOpen
  \bibfield  {author} {\bibinfo {author} {\bibfnamefont {T.~L.~M.}\ \bibnamefont {Lezama}}, \bibinfo {author} {\bibfnamefont {S.}~\bibnamefont {Bera}},\ and\ \bibinfo {author} {\bibfnamefont {J.~H.}\ \bibnamefont {Bardarson}},\ }\bibfield  {title} {\bibinfo {title} {Apparent slow dynamics in the ergodic phase of a driven many-body localized system without extensive conserved quantities},\ }\href {https://doi.org/10.1103/PhysRevB.99.161106} {\bibfield  {journal} {\bibinfo  {journal} {Phys. Rev. B}\ }\textbf {\bibinfo {volume} {99}},\ \bibinfo {pages} {161106} (\bibinfo {year} {2019})}\BibitemShut {NoStop}%
\bibitem [{\citenamefont {Sierant}\ \emph {et~al.}(2023)\citenamefont {Sierant}, \citenamefont {Lewenstein}, \citenamefont {Scardicchio},\ and\ \citenamefont {Zakrzewski}}]{SierantPRBising23}%
  \BibitemOpen
  \bibfield  {author} {\bibinfo {author} {\bibfnamefont {P.}~\bibnamefont {Sierant}}, \bibinfo {author} {\bibfnamefont {M.}~\bibnamefont {Lewenstein}}, \bibinfo {author} {\bibfnamefont {A.}~\bibnamefont {Scardicchio}},\ and\ \bibinfo {author} {\bibfnamefont {J.}~\bibnamefont {Zakrzewski}},\ }\bibfield  {title} {\bibinfo {title} {Stability of many-body localization in floquet systems},\ }\href {https://doi.org/10.1103/PhysRevB.107.115132} {\bibfield  {journal} {\bibinfo  {journal} {Phys. Rev. B}\ }\textbf {\bibinfo {volume} {107}},\ \bibinfo {pages} {115132} (\bibinfo {year} {2023})}\BibitemShut {NoStop}%
\bibitem [{\citenamefont {Kim}\ and\ \citenamefont {Huse}(2013)}]{KimPRL13}%
  \BibitemOpen
  \bibfield  {author} {\bibinfo {author} {\bibfnamefont {H.}~\bibnamefont {Kim}}\ and\ \bibinfo {author} {\bibfnamefont {D.~A.}\ \bibnamefont {Huse}},\ }\bibfield  {title} {\bibinfo {title} {Ballistic spreading of entanglement in a diffusive nonintegrable system},\ }\href {https://doi.org/10.1103/PhysRevLett.111.127205} {\bibfield  {journal} {\bibinfo  {journal} {Phys. Rev. Lett.}\ }\textbf {\bibinfo {volume} {111}},\ \bibinfo {pages} {127205} (\bibinfo {year} {2013})}\BibitemShut {NoStop}%
\bibitem [{\citenamefont {Burke}\ \emph {et~al.}(2023)\citenamefont {Burke}, \citenamefont {Nakerst},\ and\ \citenamefont {Haque}}]{BurkeHaquePRE23}%
  \BibitemOpen
  \bibfield  {author} {\bibinfo {author} {\bibfnamefont {P.~C.}\ \bibnamefont {Burke}}, \bibinfo {author} {\bibfnamefont {G.}~\bibnamefont {Nakerst}},\ and\ \bibinfo {author} {\bibfnamefont {M.}~\bibnamefont {Haque}},\ }\bibfield  {title} {\bibinfo {title} {Assigning temperatures to eigenstates},\ }\href {https://doi.org/10.1103/PhysRevE.107.024102} {\bibfield  {journal} {\bibinfo  {journal} {Phys. Rev. E}\ }\textbf {\bibinfo {volume} {107}},\ \bibinfo {pages} {024102} (\bibinfo {year} {2023})}\BibitemShut {NoStop}%
\bibitem [{\citenamefont {Lazarides}\ \emph {et~al.}(2014)\citenamefont {Lazarides}, \citenamefont {Das},\ and\ \citenamefont {Moessner}}]{LazaridesPRL14}%
  \BibitemOpen
  \bibfield  {author} {\bibinfo {author} {\bibfnamefont {A.}~\bibnamefont {Lazarides}}, \bibinfo {author} {\bibfnamefont {A.}~\bibnamefont {Das}},\ and\ \bibinfo {author} {\bibfnamefont {R.}~\bibnamefont {Moessner}},\ }\bibfield  {title} {\bibinfo {title} {Periodic thermodynamics of isolated quantum systems},\ }\href {https://doi.org/10.1103/PhysRevLett.112.150401} {\bibfield  {journal} {\bibinfo  {journal} {Phys. Rev. Lett.}\ }\textbf {\bibinfo {volume} {112}},\ \bibinfo {pages} {150401} (\bibinfo {year} {2014})}\BibitemShut {NoStop}%
\bibitem [{\citenamefont {D'Alessio}\ and\ \citenamefont {Rigol}(2014)}]{DalessioPRX14}%
  \BibitemOpen
  \bibfield  {author} {\bibinfo {author} {\bibfnamefont {L.}~\bibnamefont {D'Alessio}}\ and\ \bibinfo {author} {\bibfnamefont {M.}~\bibnamefont {Rigol}},\ }\bibfield  {title} {\bibinfo {title} {Long-time behavior of isolated periodically driven interacting lattice systems},\ }\href {https://doi.org/10.1103/PhysRevX.4.041048} {\bibfield  {journal} {\bibinfo  {journal} {Phys. Rev. X}\ }\textbf {\bibinfo {volume} {4}},\ \bibinfo {pages} {041048} (\bibinfo {year} {2014})}\BibitemShut {NoStop}%
\bibitem [{\citenamefont {Abanin}\ \emph {et~al.}(2015)\citenamefont {Abanin}, \citenamefont {De~Roeck},\ and\ \citenamefont {Huveneers}}]{AbaninPRL15}%
  \BibitemOpen
  \bibfield  {author} {\bibinfo {author} {\bibfnamefont {D.~A.}\ \bibnamefont {Abanin}}, \bibinfo {author} {\bibfnamefont {W.}~\bibnamefont {De~Roeck}},\ and\ \bibinfo {author} {\bibfnamefont {F.~m.~c.}\ \bibnamefont {Huveneers}},\ }\bibfield  {title} {\bibinfo {title} {Exponentially slow heating in periodically driven many-body systems},\ }\href {https://doi.org/10.1103/PhysRevLett.115.256803} {\bibfield  {journal} {\bibinfo  {journal} {Phys. Rev. Lett.}\ }\textbf {\bibinfo {volume} {115}},\ \bibinfo {pages} {256803} (\bibinfo {year} {2015})}\BibitemShut {NoStop}%
\bibitem [{\citenamefont {Mori}\ \emph {et~al.}(2016)\citenamefont {Mori}, \citenamefont {Kuwahara},\ and\ \citenamefont {Saito}}]{MoriPRL16}%
  \BibitemOpen
  \bibfield  {author} {\bibinfo {author} {\bibfnamefont {T.}~\bibnamefont {Mori}}, \bibinfo {author} {\bibfnamefont {T.}~\bibnamefont {Kuwahara}},\ and\ \bibinfo {author} {\bibfnamefont {K.}~\bibnamefont {Saito}},\ }\bibfield  {title} {\bibinfo {title} {Rigorous bound on energy absorption and generic relaxation in periodically driven quantum systems},\ }\href {https://doi.org/10.1103/PhysRevLett.116.120401} {\bibfield  {journal} {\bibinfo  {journal} {Phys. Rev. Lett.}\ }\textbf {\bibinfo {volume} {116}},\ \bibinfo {pages} {120401} (\bibinfo {year} {2016})}\BibitemShut {NoStop}%
\bibitem [{\citenamefont {Abanin}\ \emph {et~al.}(2017)\citenamefont {Abanin}, \citenamefont {De~Roeck}, \citenamefont {Ho},\ and\ \citenamefont {Huveneers}}]{AbaninPRB17}%
  \BibitemOpen
  \bibfield  {author} {\bibinfo {author} {\bibfnamefont {D.~A.}\ \bibnamefont {Abanin}}, \bibinfo {author} {\bibfnamefont {W.}~\bibnamefont {De~Roeck}}, \bibinfo {author} {\bibfnamefont {W.~W.}\ \bibnamefont {Ho}},\ and\ \bibinfo {author} {\bibfnamefont {F.~m.~c.}\ \bibnamefont {Huveneers}},\ }\bibfield  {title} {\bibinfo {title} {Effective hamiltonians, prethermalization, and slow energy absorption in periodically driven many-body systems},\ }\href {https://doi.org/10.1103/PhysRevB.95.014112} {\bibfield  {journal} {\bibinfo  {journal} {Phys. Rev. B}\ }\textbf {\bibinfo {volume} {95}},\ \bibinfo {pages} {014112} (\bibinfo {year} {2017})}\BibitemShut {NoStop}%
\bibitem [{\citenamefont {Mallayya}\ and\ \citenamefont {Rigol}(2019)}]{MallayyaPRL19}%
  \BibitemOpen
  \bibfield  {author} {\bibinfo {author} {\bibfnamefont {K.}~\bibnamefont {Mallayya}}\ and\ \bibinfo {author} {\bibfnamefont {M.}~\bibnamefont {Rigol}},\ }\bibfield  {title} {\bibinfo {title} {Heating rates in periodically driven strongly interacting quantum many-body systems},\ }\href {https://doi.org/10.1103/PhysRevLett.123.240603} {\bibfield  {journal} {\bibinfo  {journal} {Phys. Rev. Lett.}\ }\textbf {\bibinfo {volume} {123}},\ \bibinfo {pages} {240603} (\bibinfo {year} {2019})}\BibitemShut {NoStop}%
\bibitem [{\citenamefont {Ponte}\ \emph {et~al.}(2015)\citenamefont {Ponte}, \citenamefont {Papi\ifmmode~\acute{c}\else \'{c}\fi{}}, \citenamefont {Huveneers},\ and\ \citenamefont {Abanin}}]{PontePRL14}%
  \BibitemOpen
  \bibfield  {author} {\bibinfo {author} {\bibfnamefont {P.}~\bibnamefont {Ponte}}, \bibinfo {author} {\bibfnamefont {Z.}~\bibnamefont {Papi\ifmmode~\acute{c}\else \'{c}\fi{}}}, \bibinfo {author} {\bibfnamefont {F.~m.~c.}\ \bibnamefont {Huveneers}},\ and\ \bibinfo {author} {\bibfnamefont {D.~A.}\ \bibnamefont {Abanin}},\ }\bibfield  {title} {\bibinfo {title} {Many-body localization in periodically driven systems},\ }\href {https://doi.org/10.1103/PhysRevLett.114.140401} {\bibfield  {journal} {\bibinfo  {journal} {Phys. Rev. Lett.}\ }\textbf {\bibinfo {volume} {114}},\ \bibinfo {pages} {140401} (\bibinfo {year} {2015})}\BibitemShut {NoStop}%
\bibitem [{\citenamefont {Lazarides}\ \emph {et~al.}(2015)\citenamefont {Lazarides}, \citenamefont {Das},\ and\ \citenamefont {Moessner}}]{LazaridesMBLPRL15}%
  \BibitemOpen
  \bibfield  {author} {\bibinfo {author} {\bibfnamefont {A.}~\bibnamefont {Lazarides}}, \bibinfo {author} {\bibfnamefont {A.}~\bibnamefont {Das}},\ and\ \bibinfo {author} {\bibfnamefont {R.}~\bibnamefont {Moessner}},\ }\bibfield  {title} {\bibinfo {title} {Fate of many-body localization under periodic driving},\ }\href {https://doi.org/10.1103/PhysRevLett.115.030402} {\bibfield  {journal} {\bibinfo  {journal} {Phys. Rev. Lett.}\ }\textbf {\bibinfo {volume} {115}},\ \bibinfo {pages} {030402} (\bibinfo {year} {2015})}\BibitemShut {NoStop}%
\bibitem [{\citenamefont {Rehn}\ \emph {et~al.}(2016)\citenamefont {Rehn}, \citenamefont {Lazarides}, \citenamefont {Pollmann},\ and\ \citenamefont {Moessner}}]{RehnPRB16}%
  \BibitemOpen
  \bibfield  {author} {\bibinfo {author} {\bibfnamefont {J.}~\bibnamefont {Rehn}}, \bibinfo {author} {\bibfnamefont {A.}~\bibnamefont {Lazarides}}, \bibinfo {author} {\bibfnamefont {F.}~\bibnamefont {Pollmann}},\ and\ \bibinfo {author} {\bibfnamefont {R.}~\bibnamefont {Moessner}},\ }\bibfield  {title} {\bibinfo {title} {How periodic driving heats a disordered quantum spin chain},\ }\href {https://doi.org/10.1103/PhysRevB.94.020201} {\bibfield  {journal} {\bibinfo  {journal} {Phys. Rev. B}\ }\textbf {\bibinfo {volume} {94}},\ \bibinfo {pages} {020201} (\bibinfo {year} {2016})}\BibitemShut {NoStop}%
\bibitem [{\citenamefont {Bordia}\ \emph {et~al.}(2017)\citenamefont {Bordia}, \citenamefont {L{\"u}schen}, \citenamefont {Schneider}, \citenamefont {Knap},\ and\ \citenamefont {Bloch}}]{BordiaNatPhy2017}%
  \BibitemOpen
  \bibfield  {author} {\bibinfo {author} {\bibfnamefont {P.}~\bibnamefont {Bordia}}, \bibinfo {author} {\bibfnamefont {H.}~\bibnamefont {L{\"u}schen}}, \bibinfo {author} {\bibfnamefont {U.}~\bibnamefont {Schneider}}, \bibinfo {author} {\bibfnamefont {M.}~\bibnamefont {Knap}},\ and\ \bibinfo {author} {\bibfnamefont {I.}~\bibnamefont {Bloch}},\ }\bibfield  {title} {\bibinfo {title} {{Periodically driving a many-body localized quantum system}},\ }\href {https://doi.org/10.1038/nphys4020} {\bibfield  {journal} {\bibinfo  {journal} {Nature Physics}\ }\textbf {\bibinfo {volume} {13}},\ \bibinfo {pages} {460} (\bibinfo {year} {2017})}\BibitemShut {NoStop}%
\bibitem [{\citenamefont {Haldar}(2023)}]{Asmi23}%
  \BibitemOpen
  \bibfield  {author} {\bibinfo {author} {\bibfnamefont {A.}~\bibnamefont {Haldar}},\ }\href@noop {} {\bibinfo {title} {Slow dynamics and kohlrausch relaxation in isolated disordered many-body systems}} (\bibinfo {year} {2023}),\ \Eprint {https://arxiv.org/abs/2302.12275} {arXiv:2302.12275 [cond-mat.stat-mech]} \BibitemShut {NoStop}%
\bibitem [{\citenamefont {Ikeda}\ and\ \citenamefont {Polkovnikov}(2021)}]{TatsuhikoPRB21}%
  \BibitemOpen
  \bibfield  {author} {\bibinfo {author} {\bibfnamefont {T.~N.}\ \bibnamefont {Ikeda}}\ and\ \bibinfo {author} {\bibfnamefont {A.}~\bibnamefont {Polkovnikov}},\ }\bibfield  {title} {\bibinfo {title} {Fermi's golden rule for heating in strongly driven floquet systems},\ }\href {https://doi.org/10.1103/PhysRevB.104.134308} {\bibfield  {journal} {\bibinfo  {journal} {Phys. Rev. B}\ }\textbf {\bibinfo {volume} {104}},\ \bibinfo {pages} {134308} (\bibinfo {year} {2021})}\BibitemShut {NoStop}%
\bibitem [{\citenamefont {Rakcheev}\ and\ \citenamefont {L\"auchli}(2022)}]{RakcheevPRR22}%
  \BibitemOpen
  \bibfield  {author} {\bibinfo {author} {\bibfnamefont {A.}~\bibnamefont {Rakcheev}}\ and\ \bibinfo {author} {\bibfnamefont {A.~M.}\ \bibnamefont {L\"auchli}},\ }\bibfield  {title} {\bibinfo {title} {Estimating heating times in periodically driven quantum many-body systems via avoided crossing spectroscopy},\ }\href {https://doi.org/10.1103/PhysRevResearch.4.043174} {\bibfield  {journal} {\bibinfo  {journal} {Phys. Rev. Res.}\ }\textbf {\bibinfo {volume} {4}},\ \bibinfo {pages} {043174} (\bibinfo {year} {2022})}\BibitemShut {NoStop}%
\bibitem [{\citenamefont {Page}(1993)}]{Page93}%
  \BibitemOpen
  \bibfield  {author} {\bibinfo {author} {\bibfnamefont {D.~N.}\ \bibnamefont {Page}},\ }\bibfield  {title} {\bibinfo {title} {Average entropy of a subsystem},\ }\href {https://doi.org/10.1103/PhysRevLett.71.1291} {\bibfield  {journal} {\bibinfo  {journal} {Phys. Rev. Lett.}\ }\textbf {\bibinfo {volume} {71}},\ \bibinfo {pages} {1291} (\bibinfo {year} {1993})}\BibitemShut {NoStop}%
\bibitem [{\citenamefont {Lakshminarayan}(2001)}]{ArulPRE01}%
  \BibitemOpen
  \bibfield  {author} {\bibinfo {author} {\bibfnamefont {A.}~\bibnamefont {Lakshminarayan}},\ }\bibfield  {title} {\bibinfo {title} {Entangling power of quantized chaotic systems},\ }\href {https://doi.org/10.1103/PhysRevE.64.036207} {\bibfield  {journal} {\bibinfo  {journal} {Phys. Rev. E}\ }\textbf {\bibinfo {volume} {64}},\ \bibinfo {pages} {036207} (\bibinfo {year} {2001})}\BibitemShut {NoStop}%
\bibitem [{\citenamefont {De~Tomasi}\ and\ \citenamefont {Khaymovich}(2020)}]{DetomasiPRL20}%
  \BibitemOpen
  \bibfield  {author} {\bibinfo {author} {\bibfnamefont {G.}~\bibnamefont {De~Tomasi}}\ and\ \bibinfo {author} {\bibfnamefont {I.~M.}\ \bibnamefont {Khaymovich}},\ }\bibfield  {title} {\bibinfo {title} {Multifractality meets entanglement: Relation for nonergodic extended states},\ }\href {https://doi.org/10.1103/PhysRevLett.124.200602} {\bibfield  {journal} {\bibinfo  {journal} {Phys. Rev. Lett.}\ }\textbf {\bibinfo {volume} {124}},\ \bibinfo {pages} {200602} (\bibinfo {year} {2020})}\BibitemShut {NoStop}%
\bibitem [{\citenamefont {De~Roeck}\ and\ \citenamefont {Huveneers}(2017)}]{deroeckPRB17}%
  \BibitemOpen
  \bibfield  {author} {\bibinfo {author} {\bibfnamefont {W.}~\bibnamefont {De~Roeck}}\ and\ \bibinfo {author} {\bibfnamefont {F.~m.~c.}\ \bibnamefont {Huveneers}},\ }\bibfield  {title} {\bibinfo {title} {Stability and instability towards delocalization in many-body localization systems},\ }\href {https://doi.org/10.1103/PhysRevB.95.155129} {\bibfield  {journal} {\bibinfo  {journal} {Phys. Rev. B}\ }\textbf {\bibinfo {volume} {95}},\ \bibinfo {pages} {155129} (\bibinfo {year} {2017})}\BibitemShut {NoStop}%
\bibitem [{\citenamefont {Thiery}\ \emph {et~al.}(2018)\citenamefont {Thiery}, \citenamefont {Huveneers}, \citenamefont {M\"uller},\ and\ \citenamefont {De~Roeck}}]{Thiery2017}%
  \BibitemOpen
  \bibfield  {author} {\bibinfo {author} {\bibfnamefont {T.}~\bibnamefont {Thiery}}, \bibinfo {author} {\bibfnamefont {F.}~\bibnamefont {Huveneers}}, \bibinfo {author} {\bibfnamefont {M.}~\bibnamefont {M\"uller}},\ and\ \bibinfo {author} {\bibfnamefont {W.}~\bibnamefont {De~Roeck}},\ }\bibfield  {title} {\bibinfo {title} {Many-body delocalization as a quantum avalanche},\ }\href {https://doi.org/10.1103/PhysRevLett.121.140601} {\bibfield  {journal} {\bibinfo  {journal} {Phys. Rev. Lett.}\ }\textbf {\bibinfo {volume} {121}},\ \bibinfo {pages} {140601} (\bibinfo {year} {2018})}\BibitemShut {NoStop}%
\bibitem [{\citenamefont {Gopalakrishnan}\ \emph {et~al.}(2015)\citenamefont {Gopalakrishnan}, \citenamefont {M\"uller}, \citenamefont {Khemani}, \citenamefont {Knap}, \citenamefont {Demler},\ and\ \citenamefont {Huse}}]{GopalakrishnanPRB15}%
  \BibitemOpen
  \bibfield  {author} {\bibinfo {author} {\bibfnamefont {S.}~\bibnamefont {Gopalakrishnan}}, \bibinfo {author} {\bibfnamefont {M.}~\bibnamefont {M\"uller}}, \bibinfo {author} {\bibfnamefont {V.}~\bibnamefont {Khemani}}, \bibinfo {author} {\bibfnamefont {M.}~\bibnamefont {Knap}}, \bibinfo {author} {\bibfnamefont {E.}~\bibnamefont {Demler}},\ and\ \bibinfo {author} {\bibfnamefont {D.~A.}\ \bibnamefont {Huse}},\ }\bibfield  {title} {\bibinfo {title} {Low-frequency conductivity in many-body localized systems},\ }\href {https://doi.org/10.1103/PhysRevB.92.104202} {\bibfield  {journal} {\bibinfo  {journal} {Phys. Rev. B}\ }\textbf {\bibinfo {volume} {92}},\ \bibinfo {pages} {104202} (\bibinfo {year} {2015})}\BibitemShut {NoStop}%
\bibitem [{\citenamefont {Garratt}\ \emph {et~al.}(2021)\citenamefont {Garratt}, \citenamefont {Roy},\ and\ \citenamefont {Chalker}}]{Garratt2021}%
  \BibitemOpen
  \bibfield  {author} {\bibinfo {author} {\bibfnamefont {S.~J.}\ \bibnamefont {Garratt}}, \bibinfo {author} {\bibfnamefont {S.}~\bibnamefont {Roy}},\ and\ \bibinfo {author} {\bibfnamefont {J.~T.}\ \bibnamefont {Chalker}},\ }\bibfield  {title} {\bibinfo {title} {Local resonances and parametric level dynamics in the many-body localized phase},\ }\href {https://doi.org/10.1103/PhysRevB.104.184203} {\bibfield  {journal} {\bibinfo  {journal} {Phys. Rev. B}\ }\textbf {\bibinfo {volume} {104}},\ \bibinfo {pages} {184203} (\bibinfo {year} {2021})}\BibitemShut {NoStop}%
\bibitem [{\citenamefont {Garratt}\ and\ \citenamefont {Roy}(2022)}]{GarrattPRB22}%
  \BibitemOpen
  \bibfield  {author} {\bibinfo {author} {\bibfnamefont {S.~J.}\ \bibnamefont {Garratt}}\ and\ \bibinfo {author} {\bibfnamefont {S.}~\bibnamefont {Roy}},\ }\bibfield  {title} {\bibinfo {title} {Resonant energy scales and local observables in the many-body localized phase},\ }\href {https://doi.org/10.1103/PhysRevB.106.054309} {\bibfield  {journal} {\bibinfo  {journal} {Phys. Rev. B}\ }\textbf {\bibinfo {volume} {106}},\ \bibinfo {pages} {054309} (\bibinfo {year} {2022})}\BibitemShut {NoStop}%
\bibitem [{\citenamefont {L{\'e}onard}\ \emph {et~al.}(2023)\citenamefont {L{\'e}onard}, \citenamefont {Kim}, \citenamefont {Rispoli}, \citenamefont {Lukin}, \citenamefont {Schittko}, \citenamefont {Kwan}, \citenamefont {Demler}, \citenamefont {Sels},\ and\ \citenamefont {Greiner}}]{LeonardNatPhys23}%
  \BibitemOpen
  \bibfield  {author} {\bibinfo {author} {\bibfnamefont {J.}~\bibnamefont {L{\'e}onard}}, \bibinfo {author} {\bibfnamefont {S.}~\bibnamefont {Kim}}, \bibinfo {author} {\bibfnamefont {M.}~\bibnamefont {Rispoli}}, \bibinfo {author} {\bibfnamefont {A.}~\bibnamefont {Lukin}}, \bibinfo {author} {\bibfnamefont {R.}~\bibnamefont {Schittko}}, \bibinfo {author} {\bibfnamefont {J.}~\bibnamefont {Kwan}}, \bibinfo {author} {\bibfnamefont {E.}~\bibnamefont {Demler}}, \bibinfo {author} {\bibfnamefont {D.}~\bibnamefont {Sels}},\ and\ \bibinfo {author} {\bibfnamefont {M.}~\bibnamefont {Greiner}},\ }\bibfield  {title} {\bibinfo {title} {Probing the onset of quantum avalanches in a many-body localized system},\ }\href {https://doi.org/10.1038/s41567-022-01887-3} {\bibfield  {journal} {\bibinfo  {journal} {Nature Physics}\ }\textbf {\bibinfo {volume} {19}},\ \bibinfo {pages} {481} (\bibinfo {year} {2023})}\BibitemShut {NoStop}%
\bibitem [{\citenamefont {Bera}\ \emph {et~al.}(2017)\citenamefont {Bera}, \citenamefont {De~Tomasi}, \citenamefont {Weiner},\ and\ \citenamefont {Evers}}]{Bera2017}%
  \BibitemOpen
  \bibfield  {author} {\bibinfo {author} {\bibfnamefont {S.}~\bibnamefont {Bera}}, \bibinfo {author} {\bibfnamefont {G.}~\bibnamefont {De~Tomasi}}, \bibinfo {author} {\bibfnamefont {F.}~\bibnamefont {Weiner}},\ and\ \bibinfo {author} {\bibfnamefont {F.}~\bibnamefont {Evers}},\ }\bibfield  {title} {\bibinfo {title} {Density propagator for many-body localization: Finite-size effects, transient subdiffusion, and exponential decay},\ }\href {https://doi.org/10.1103/PhysRevLett.118.196801} {\bibfield  {journal} {\bibinfo  {journal} {Phys. Rev. Lett.}\ }\textbf {\bibinfo {volume} {118}},\ \bibinfo {pages} {196801} (\bibinfo {year} {2017})}\BibitemShut {NoStop}%
\bibitem [{\citenamefont {Chanda}\ \emph {et~al.}(2020)\citenamefont {Chanda}, \citenamefont {Sierant},\ and\ \citenamefont {Zakrzewski}}]{ChandaPRB20}%
  \BibitemOpen
  \bibfield  {author} {\bibinfo {author} {\bibfnamefont {T.}~\bibnamefont {Chanda}}, \bibinfo {author} {\bibfnamefont {P.}~\bibnamefont {Sierant}},\ and\ \bibinfo {author} {\bibfnamefont {J.}~\bibnamefont {Zakrzewski}},\ }\bibfield  {title} {\bibinfo {title} {Time dynamics with matrix product states: Many-body localization transition of large systems revisited},\ }\href {https://doi.org/10.1103/PhysRevB.101.035148} {\bibfield  {journal} {\bibinfo  {journal} {Phys. Rev. B}\ }\textbf {\bibinfo {volume} {101}},\ \bibinfo {pages} {035148} (\bibinfo {year} {2020})}\BibitemShut {NoStop}%
\bibitem [{\citenamefont {Sierant}\ \emph {et~al.}(2020)\citenamefont {Sierant}, \citenamefont {Lewenstein},\ and\ \citenamefont {Zakrzewski}}]{SierantLargeWc20}%
  \BibitemOpen
  \bibfield  {author} {\bibinfo {author} {\bibfnamefont {P.}~\bibnamefont {Sierant}}, \bibinfo {author} {\bibfnamefont {M.}~\bibnamefont {Lewenstein}},\ and\ \bibinfo {author} {\bibfnamefont {J.}~\bibnamefont {Zakrzewski}},\ }\bibfield  {title} {\bibinfo {title} {Polynomially filtered exact diagonalization approach to many-body localization},\ }\href {https://doi.org/10.1103/PhysRevLett.125.156601} {\bibfield  {journal} {\bibinfo  {journal} {Phys. Rev. Lett.}\ }\textbf {\bibinfo {volume} {125}},\ \bibinfo {pages} {156601} (\bibinfo {year} {2020})}\BibitemShut {NoStop}%
\bibitem [{\citenamefont {Doggen}\ \emph {et~al.}(2021)\citenamefont {Doggen}, \citenamefont {Gornyi}, \citenamefont {Mirlin},\ and\ \citenamefont {Polyakov}}]{DoggenRevAnnPhy21}%
  \BibitemOpen
  \bibfield  {author} {\bibinfo {author} {\bibfnamefont {E.~V.}\ \bibnamefont {Doggen}}, \bibinfo {author} {\bibfnamefont {I.~V.}\ \bibnamefont {Gornyi}}, \bibinfo {author} {\bibfnamefont {A.~D.}\ \bibnamefont {Mirlin}},\ and\ \bibinfo {author} {\bibfnamefont {D.~G.}\ \bibnamefont {Polyakov}},\ }\bibfield  {title} {\bibinfo {title} {Many-body localization in large systems: Matrix-product-state approach},\ }\href {https://doi.org/https://doi.org/10.1016/j.aop.2021.168437} {\bibfield  {journal} {\bibinfo  {journal} {Ann. Phys.}\ }\textbf {\bibinfo {volume} {435}},\ \bibinfo {pages} {168437} (\bibinfo {year} {2021})}\BibitemShut {NoStop}%
\end{thebibliography}%

\end{document}